\documentclass{mn2e}
\usepackage{epsfig}
\usepackage{times}

\title{{\it XMM-Newton} observations of the hot-gas atmospheres of 3C 66B and 3C 449} 
\author[J. H. Croston et al.]
       {J. H. Croston, \thanks{Email: Judith.Croston@bris.ac.uk} 
	M. J. Hardcastle, 
	M. Birkinshaw, 
	and D. M. Worrall \\
        H. H. Wills Physics Laboratory, University of Bristol, Tyndall Avenue, Bristol BS8 1TL }

\date{September 2002}

\pagerange{\pageref{firstpage}--\pageref{lastpage}}
\pubyear{2003}
\begin{document}

\maketitle

\label{firstpage}

\begin{abstract}

We present new {\it XMM-Newton} observations of the hot-gas environments of two low-power twin-jet radio galaxies, 3C 66B and 3C 449, showing direct evidence for the interactions between X-ray-emitting gas and radio plasma that are thought to determine the large-scale radio structure of these sources. The temperatures that we measure for the two environments are significantly higher than those predicted by standard luminosity-temperature relations for clusters and groups. We show that luminosity-temperature relations for radio-quiet and radio-loud X-ray groups differ, in the sense that radio-source heating may operate in most groups containing radio galaxies. If the radio lobes are expanding subsonically, we find minimum ages of $3 \times 10^{8}$ years for 3C 66B, and $5 \times 10^{8}$ years for 3C 449, older than the values obtained from spectral ageing, which would give the radio source sufficient time to heat the groups to the observed temperatures for plausible values of the jet power. The external pressures in the atmospheres of both radio galaxies are an order of magnitude higher than equipartition estimates of their radio-lobe pressures, confirming that the radio lobes are either out of equipartition or require a pressure contribution from non-radiating particles. Constraints from the level of X-ray emission we measure from the radio lobes allow us to conclude that a departure from equipartition must be in the direction of magnetic domination, and that the most plausible candidates for a particle contribution to lobe pressure are relativistic protons, an additional population of low-energy electrons, or entrained and heated thermal material.

\end{abstract}


\section{Introduction}

The radio lobes of twin-jet low-power (FR-I) radio galaxies exhibit a wide range of morphologies, ranging from near symmetry to highly irregular and distorted structure (see the 3CRR Atlas -- http://www.jb.man.ac.uk/atlas/ -- for examples). Their large-scale structure is thought to be determined at least in part by interactions with the surrounding hot gas.  Although earlier evidence for such interactions exists (B\"ohringer et al. 1993; Hardcastle, Worrall \& Birkinshaw 1998), {\it Chandra} and {\it XMM-Newton} now provide the required sensitivity and spatial and spectral resolution for more detailed investigation, and several examples of `holes' in intracluster gas, with a clear correspondence to both the position and shape of radio-galaxy lobes, have been observed on scales of a few kpc with {\it Chandra } [e.g. Hydra A, McNamara et al. (2000); Perseus A, Fabian et al. (2000)]. Observations of the hot-gas atmospheres of radio galaxies have the potential to answer many questions relating to the energetics of radio jets and lobes. One particular problem is how to reconcile minimum-energy radio-lobe pressures with the systematically higher values determined for the external medium from X-ray measurements (e.g. Morganti et al. 1988; Worrall \& Birkinshaw 2000). Study of the physical conditions and dynamics of the X-ray-emitting gas with good spatial resolution are important in clarifying this situation. 

The X-ray properties of gas in clusters are well studied (e.g. Sarazin 1986; Fabian 1994), and Forman, Jones \& Tucker (1985) first showed evidence for the ubiquity of galaxy-scale hot-gas environments in ellipticals.  An analysis of the spectral and spatial distribution of the X-ray gas atmospheres of B2 radio galaxies was carried out by Worrall \& Birkinshaw (2000), who found large variation in the linear size of the atmospheres, with no correlation between radio-galaxy size and the scale or density of the atmosphere; later work (Worrall \& Birkinshaw 2001) found the same sample to have properties consistent with the luminosity-temperature relation for groups (Helsdon \& Ponman 2000). Recent work on interactions between radio sources and X-ray gas has focused on the relationship between AGN and cooling-flow clusters. The details of this connection have become a subject of importance because of recent X-ray spectroscopy that conflicts with the standard cooling-flow models (Peterson et al. 2003; Fabian et al. 2002). Earlier work had already suggested the presence of a feedback mechanism to explain what happens to the large quantities of cooled material in the centre of cooling clusters (Tabor \& Binney 1993). Simulations of the evolution of radio plasma bubbles in cluster gas (Churazov et al. 2002; Br\"uggen \& Kaiser 2002) show that heating can be provided in this way; however, the details of possible energy transfer between AGN and intracluster gas are still poorly understood. Although heating in cooling-flow regions of clusters is an important topic of study, the same heating processes will also occur in ordinary clusters and groups. Larger and more powerful FR-I sources are usually found in groups, where their presence is likely to have a more dramatic impact on the surrounding gas.

In the more powerful FR-II radio galaxies, shock heating due to the supersonic expansion of radio lobes is an obvious candidate for a means of transferring energy from AGN to their environments. The first direct evidence for local heating of an X-ray atmosphere by radio-lobe expansion was found in Centaurus A (Kraft et al. 2003): a rim of brighter overpressured gas capping an inner lobe has a temperature of 2.9 keV, while the surrounding ISM is ten times cooler. However, the heating process in Cen A is through shocks, as this inner lobe is thought to be expanding supersonically. The nature of interactions between typical FR-I radio galaxies and X-ray-emitting gas is somewhat different. The lobes of low-power radio galaxies are believed to be expanding subsonically on large scales, so that strong shocks would not be expected. Nevertheless, a large proportion of the radio galaxy's overall energy goes into displacing gas to inflate the lobes (B\"ohringer et al. 2002), and this energy must be dissipated into the atmosphere. This could lead to local heating at the lobe edges, or low-frequency sound waves may cause more general (and less easily detectable) heating. The presence or absence of these effects can now be investigated with {\it XMM-Newton}. 

\begin{figure*}
\begin{center}
\centerline{\hbox{
\epsfig{figure=66b_cont.ps,width=8cm}
\epsfig{figure=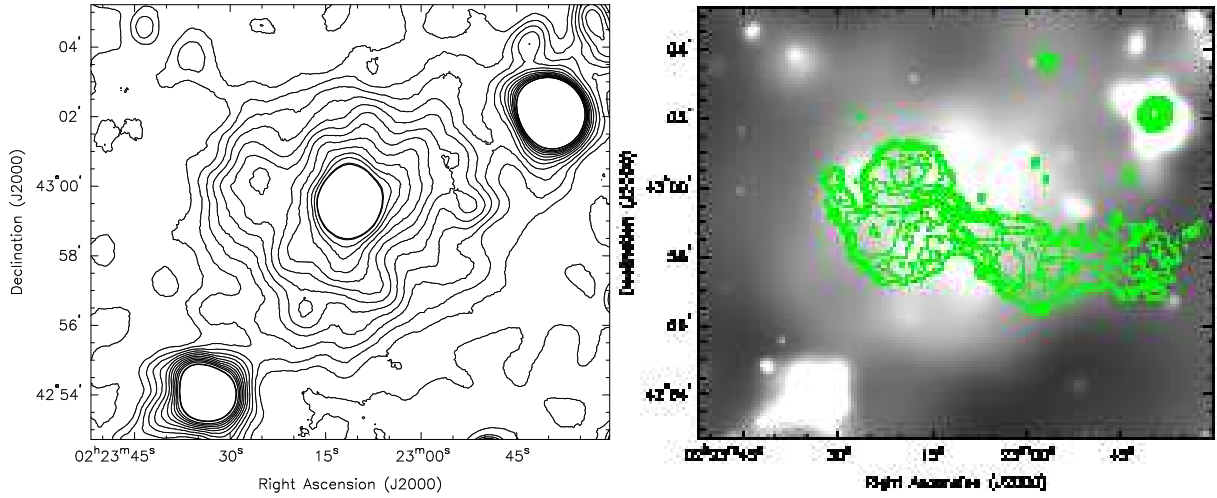,width=8cm}}}
\caption{Images of the 0.3-5.0 keV X-ray emission surrounding 3C 66B. The left-hand contour plot is of the point-source subtracted, smoothed ($\sigma$=22 arcsec), vignetting-corrected combined MOS1, MOS2 and pn data, contour levels are 0.13,0.2,0.3,0.4,0.45,0.5,0.55,0.65,0.7,0.75,0.8,0.85,0.9,0.95,1 counts/pix; the right hand image shows the point-source subtracted, adaptively smoothed, vignetting corrected, combined MOS1, MOS2 and pn data, with 1.4-GHz radio contours overlaid (from a VLA map of Hardcastle et al. 1996). Radio contour levels are $\sqrt{2}$,2,4...512 $\times$ 1.0 $\times$ 10$^{-3}$ Jy/beam. The X-ray image scale is 2 arcsec/pix.}

\label{66bimages}
\end{center}
\end{figure*}

\begin{figure*}
\begin{center}
\centerline{\hbox{
\epsfig{figure=449_cont.ps,width=8.0cm}
\epsfig{figure=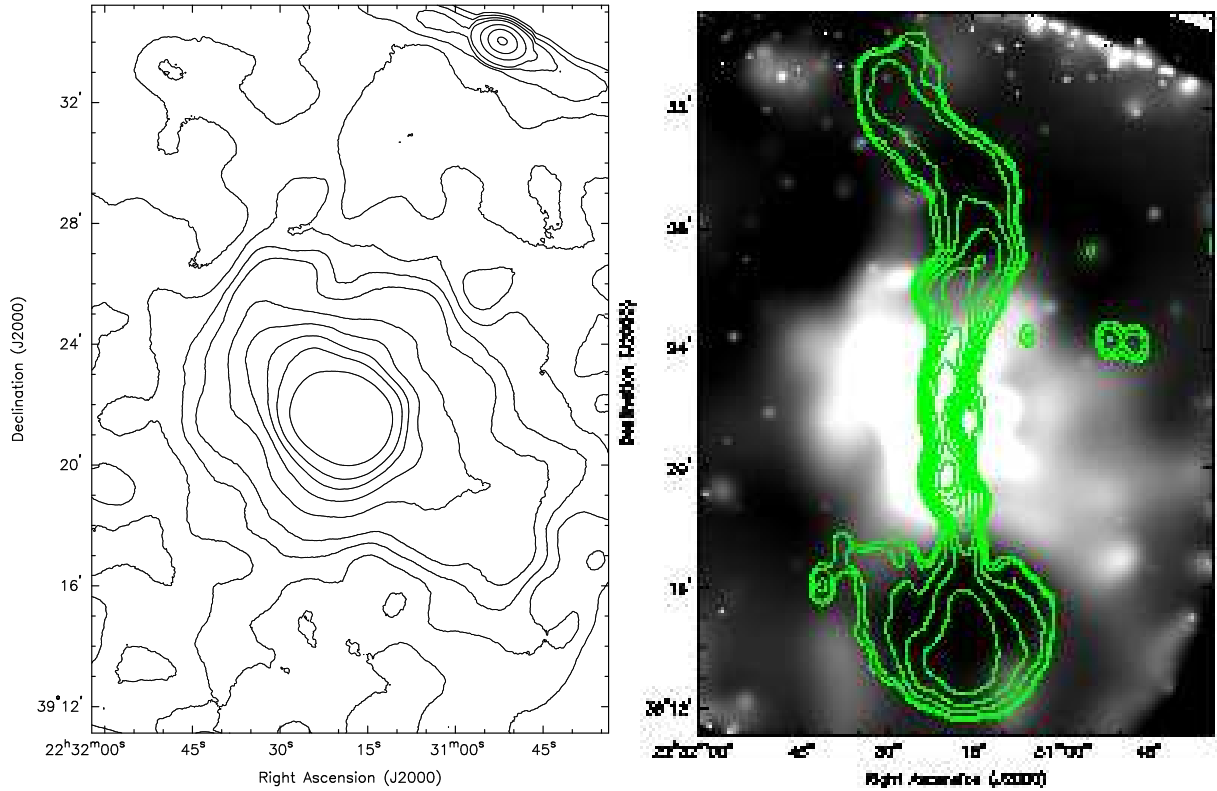,width=8.0cm}}}
\caption{Images of the 0.3-5.0 keV X-ray emission surrounding 3C 449. The left-hand contour plot is of the smoothed ($\sigma$ = 32 arcsec) vignetting-corrected combined data; contour levels are 0.25, 0.3, 0.35, 0.4, 0.5, 0.6 , 0.7, 0.8, 1 count/pix. The right hand image shows the adaptively-smoothed, vignetting corrected, combined data, with 604 MHz radio contours overlaid (from a WSRT map taken from the 3CRR atlas -- http://www.jb.man.ac.uk/atlas/object/3C449.html), radio contour levels are $\sqrt{2}$,2,4,..512 $\times 2.0 \times 10^{-3}$ Jy/beam. The X-ray image scale is 2 arcsec/pix.}
\label{449images}
\end{center}
\end{figure*}

This paper presents the analysis of new {\it XMM-Newton} observations of the environments of two low-power radio galaxies, 3C 66B and 3C 449, at $z = 0.0215$ and $z=0.0171$, respectively. These observations were obtained to look for evidence of interactions between the radio source and X-ray-emitting gas. Both sources have been extensively studied in the radio, and were chosen due to their interesting, asymmetric  radio-lobe morphologies. 3C 66B has a well-studied optical and X-ray jet (Butcher et al. 1980; Hardcastle et al. 2001). Its radio lobes are of different shapes (Fig.~\ref{66bimages}, right): the eastern lobe has sharp, well-defined edges, and a `wall' where the jet appears to stop abruptly, with some lobe emission extending further out below the jet, whereas the western lobe extends into the noise. 3C 449 has similarly interesting radio morphology, shown in the detailed multi-frequency VLA observations of Feretti et al. (1999). It has a rounded southern lobe and narrower plume in the north (Fig.~\ref{449images}, right). {\it ROSAT} observations (Hardcastle et al. 1998) showed the southern lobe to be embedded in a rim of hot gas, with a deficit at the position of the lobe suggesting interactions with the surrounding material, whereas the northern jet is free of X-ray-emitting material.  

In this paper we investigate the relationship between the X-ray-emitting gas and radio plasma, present new images of the extended emission and detailed mapping of the physical conditions in the atmospheres, and discuss the possibility that the radio galaxies are heating their environments at a detectable level. 

We use a cosmology with H$_{0}$ = 70 km s$^{-1}$ Mpc$^{-1}$, $\Omega_{M}$ = 0.3, and $\Omega_{\Lambda}$ = 0.7 throughout, which gives a scale of 26.1 kpc arcmin$^{-1}$ at the distance of 3C 66B, and a scale of 20.9 kpc arcmin$^{-1}$ at the distance of 3C 449.

\section{Data reduction and analysis}

We present results for the {\it XMM-Newton} EPIC MOS1, MOS2 and pn observations of 3C 66B and 3C 449. The duration of the 3C 66B observation was 22227 s for MOS1 and MOS2 and 19949 s for pn. For 3C 449 the duration was 21109 s for MOS1, 21117 s for MOS2, and 18498 s for pn. 

The data were reduced using the {\it XMM-Newton} Scientific Analysis Software (SAS) package, and filtered according to the methods described in the XMM-SAS Handbook. Filtering for low background was done by excluding time intervals with count rates above a threshold value obtained by studying a histogram of the count rate above 10 keV; the data were also filtered using the standard flags \#XMMEA\_EM/P, and filtered for patterns less than or equal to 12 for the MOS cameras, and less than or equal to 4 for the pn data, as suggested in the Handbook. We additionally filtered for bad columns and rows. The same filtering was applied to the background products used in our analysis (described below). Filtering resulted in good data of duration 17915 s, 17862 s, and 14692 s (MOS1, MOS2, and pn) for 3C 66B, and 20865 s, 20837 s, and 16708 s (MOS1, MOS2, and pn) for 3C 449.

Images and spectra were produced from the reduced data with the {\sc sas} task {\it evselect}, using the task {\it evigweight} to apply a vignetting correction (as described by Arnaud et al. 2002). To obtain the best possible images of the extended emission, several techniques were required. An interpolation routine was written that uses Poisson statistics to fill in the gaps due to chip boundaries and bad rows or columns, which would otherwise interfere with smoothing. The images were then adaptively smoothed using the {\sc ciao} task {\it csmooth} in order to facilitate identification of point sources. Contaminating point sources were then removed from the original unsmoothed images, using the {\sc ciao} task {\it dmfilth}, which interpolates over point source regions using the Poisson distribution of nearby background regions. The resulting images were then smoothed using Gaussian kernels to diplay the extended emission, and also adaptively smoothed to make more evident any compact structure in the extended emission.

\begin{figure*}
\begin{center}
\centerline{\vbox{\hbox{
\epsfig{figure=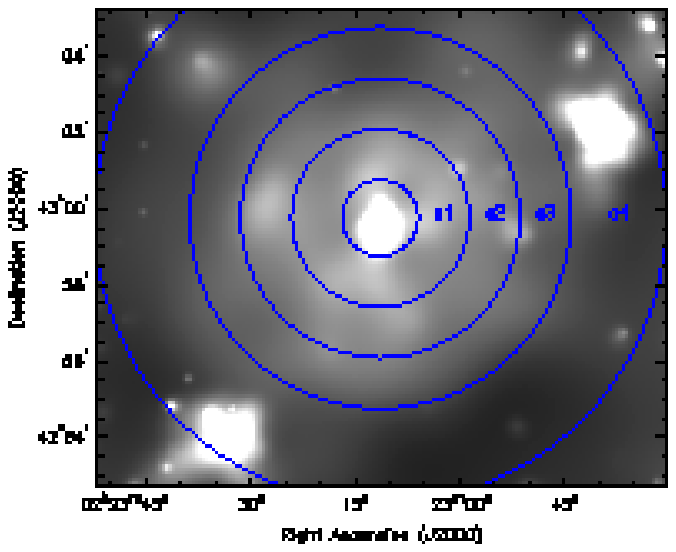,width=8.0cm}
\epsfig{figure=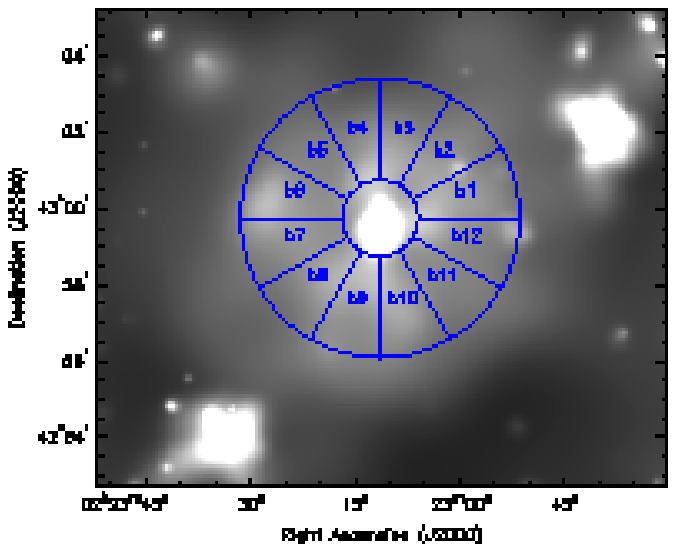,width=8.0cm}}
\hbox{
\epsfig{figure=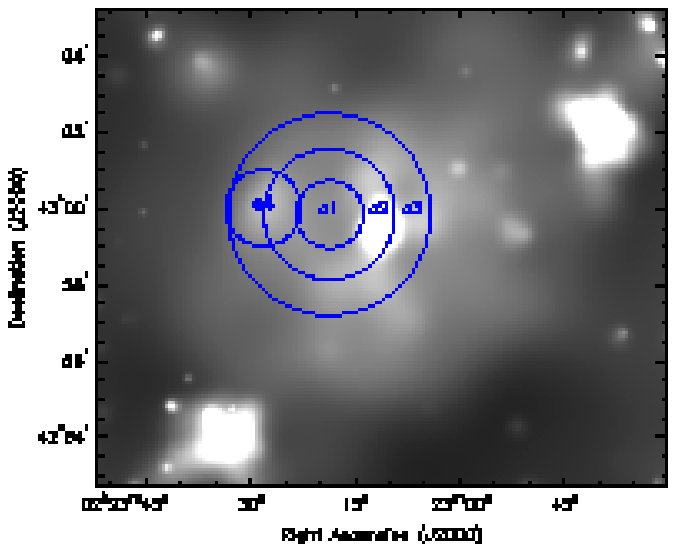,width=8.0cm}
\epsfig{figure=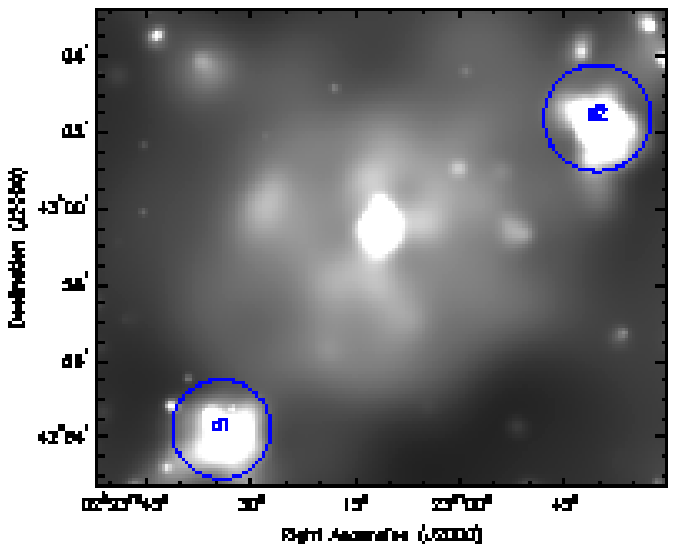,width=8.0cm}}}}
\caption{Spectral extraction regions for 3C 66B; labels indicate the regions listed in Table~\ref{66bspec}, except d1 and d2 which are the regions used for analysis of the nearby cluster and 3C 66A respectively. Region a4 excludes d1 and d2 from the extraction region, and regions c1,2 and 3 exclude a circle of radius 60 arcsec centred on the nucleus.}
\label{66bregs}
\end{center}
\end{figure*}

\begin{figure*}
\begin{center}
\centerline{\vbox{\hbox{
\epsfig{figure=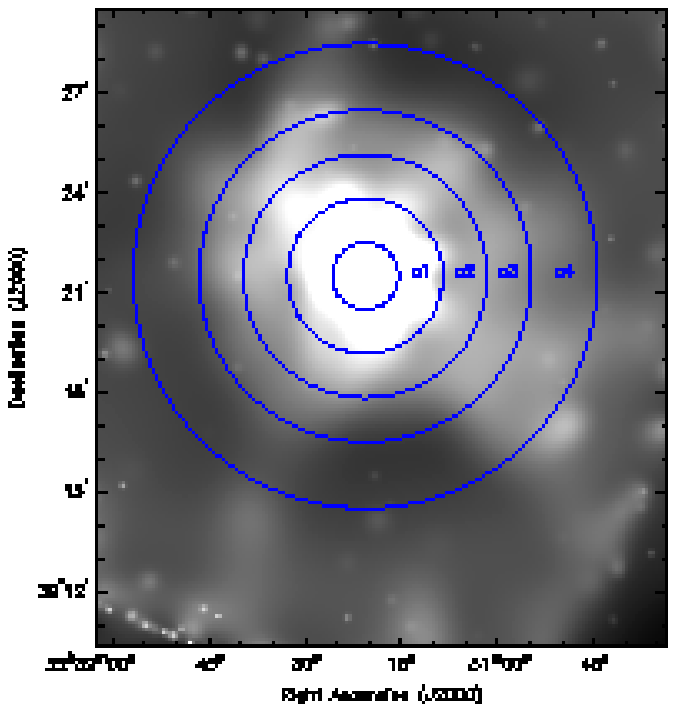,width=8.0cm}
\epsfig{figure=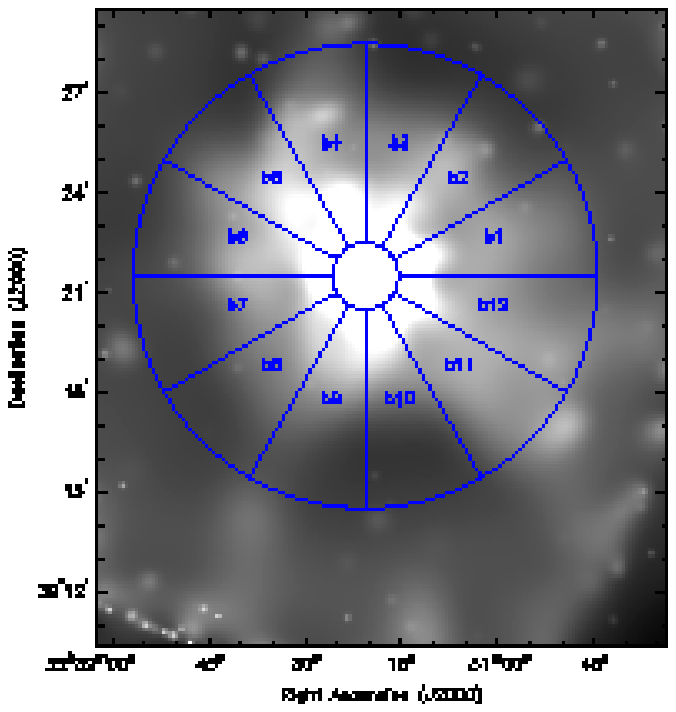,width=8.0cm}}
\epsfig{figure=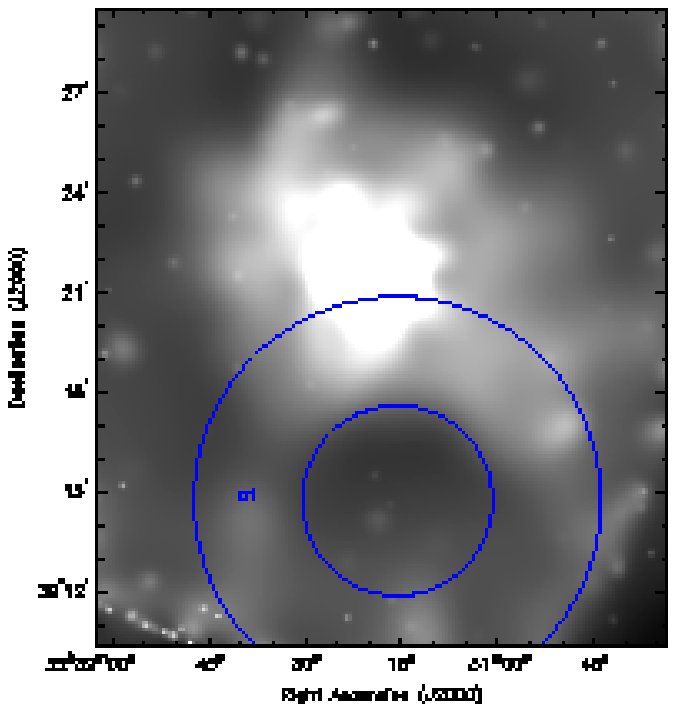,width=8.0cm}}}
\caption{Spectral extraction regions for 3C 449: labels indicate the regions listed in Table~\ref{449spec}. Region c1 excludes a circle of radius 60 arcsec centred on the nucleus.}
\label{449regs}
\end{center}
\end{figure*}

For each dataset, a series of spectral extraction regions were chosen to look for AGN components, radial and angular temperature gradients, and temperature changes related to features in the X-ray emission or the radio-source morphology. Figs.~\ref{66bregs} and ~\ref{449regs} illustrate the spectral regions for 3C 66B and 3C 449 respectively. The extracted spectra were vignetting-corrected, as described above, so that on-axis response files could be used. Response files m1\_r6\_all\_15.rmf, m2\_r6\_all\_15.rmf, and epn\_ff20\_sdY9.rmf were obtained from the {\it XMM-Newton} ftp site; appropriate ancillary response files were generated using the {\sc sas} task {\it arfgen}.

For spectral analysis of extended emission, it is essential to obtain an accurate description of the background as a function of position and of energy. Detailed analysis of the {\it XMM-Newton} background by Lumb et al. (2002) has led to background template files useful for analysis of extended emission. Techniques have been developed to correct for vignetted and non-vignetted components of the background and for variations in the soft X-ray component in different directions. Read (2003) has undertaken a detailed study of the components of the {\it XMM-Newton} background, and has created background event files and maps for different instrument, mode and filter combinations. As our observations used the medium filter, whereas the Lumb templates (available from the {\it XMM-Newton} website\footnote{http://xmm.vilspa.esa.es/}) are only available for the thin filter, we decided that Read's event files\footnote{available from http://www.sr.bham.ac.uk/xmm3/BGproducts.html} were best suited to our analysis. We filtered these files in the same way as our datasets, and applied the same vignetting correction.

The main components of the {\it XMM-Newton} background are a particle component from non-X-ray events in the detector, which dominates at hard energies, and Galactic and extragalactic X-ray background, dominating at the softer end of the spectrum. The particle background level is known to vary in intensity during and between observations, and so in the case where a local estimate of background can be obtained from the source data, a scaling factor can be used to account for the difference in particle background level between source and background observations. Unfortunately this is not the case for either of our datasets. In both cases the extended emission from our objects covers most of the field of view, so that it was not possible to obtain an estimate of the local background level. This is particularly problematic in the case of 3C 449, where a large molecular cloud in our Galaxy lies along the line-of-sight, so that the absorbing column density is high. In both cases scaling the background spectra would potentially introduce more uncertainty into an already poorly determined component of the background.

We determined the necessary scaling factors between the particle levels in our observations and the background files by using counts measured in the out of field-of-view regions (due entirely to non-X-ray events) in our energy range of interest, 0.3 -- 5.0 keV, as we found that there was significant spectral variation in particle background level, but little spatial variation across the detector (before applying the vignetting correction). To determine the amount of error introduced into our spectral analysis by our background, we used these scaling factors to determine the expected contribution of particle background to the count level in the region of lowest surface brightness for each dataset (an outer annulus), where the background contribution would be most important. We found maximum offsets in 0.3-5.0 keV counts of 0.5, 3 and 7 per cent (MOS1, MOS2, and pn) for 3C 66B, and 9, 8 and 16 per cent for 3C 449, between images with scaled and unscaled background subtraction. As this uncertainty in counts is not negligible, particularly in the case of 3C 449, we compared the {\it mekal} model fits obtained using scaled and unscaled background spectra, and found that the best-fitting temperatures did not differ significantly for the worst case scenarios for either 3C 66B or 3C 449. There was a small systematic difference in the best-fitting abundances for the pn spectrum of 3C 449; however, the unscaled value is more physically plausible and in better agreement with the other two cameras. We concluded, therefore, that the effect of background scaling on the measured physical parameters is largely insignificant, particularly compared to the unknown error due to the difference in soft Galactic background between our source and background datasets, and so we used unscaled background data in our {\sc xspec} spectral fitting. 

In order to study the spatial distribution of the X-ray-emitting gas, we obtained radial surface-brightness profiles of the two objects. The profiles were fitted with convolved point-source- and $\beta$-models following the method of Birkinshaw and Worrall (1993), individually for each camera, so as to correctly model the different camera responses. The PSF components were modelled using the analytical description provided in the document XMM-SOC-CAL-TN-0022 obtained from the XMM website. The background determination for this stage of the analysis was done in a different way to that of the spectral analysis, as a small error in the chosen background level could have a significant effect on the shape of the radial profile; in particular, the contribution of the particle background is a function of radius, since the vignetting correction is incorrectly applied to the particle-background component in our approach. Therefore, for this part of the analysis, we used a double-subtraction technique. We determined background levels in each annulus from the Read background files; these were then scaled by the factors given above and subtracted from the source counts in each bin. We then subtracted the remaining count density  in an outermost annulus (already background corrected in the first stage of background subtraction) from all the bins, so as to correct for any differences in diffuse Galactic background level between the source and background datasets.

Once temperatures had been determined from spectral fitting, and the surface-brightness profiles modelled as described above, this information was used to determine densities and pressures in the X-ray-emitting gas following the method of Birkinshaw and Worrall (1993), to compare these with the properties of the radio lobes.

\section{Results}

\subsection{3C 66B}

\begin{table*}
\centering
\label{spec}
\begin{tabular}{@{}llllll@{}}
\hline
Region & Net counts & Temp (keV) & Abundance & $\chi^{2}/n$ (d.o.f.) \\
\hline
tot & 37593 & 1.73$^{+0.03}_{-0.04}$ & 0.28$\pm$0.03 & 1.14 (718) \\
a1 & 7029 & 1.99$\pm$0.11 & 0.45$^{+0.09}_{-0.08}$ & 1.02 (269) \\
a2 & 9070 & 2.04$^{+0.10}_{-0.12}$ & 0.40$^{+0.08}_{-0.07}$ & 1.15 (330) \\
a3 & 8748 & 1.80$^{+0.09}_{-0.06}$ & 0.31$\pm$0.06 & 1.17 (311) \\
a4 & 12613 & 1.44$^{+0.05}_{-0.02}$ & 0.18$^{+0.03}_{-0.02}$ & 1.02 (395) \\
b1 & 1235 & 1.97$^{+0.28}_{-0.19}$ & 0.64$^{+0.39}_{-0.24}$ & 0.90 (54) \\
b2 & 1392 & 2.25$^{+0.45}_{-0.25}$ & 0.67$^{+0.47}_{-0.23}$ & 0.97 (61) \\
b3 & 1359 & 2.07$^{+0.24}_{-0.25}$ & 0.64$^{+0.35}_{-0.22}$ & 0.93 (59) \\
b4 & 1235 & 2.55$^{+0.56}_{-0.43}$ & 0.50$^{+0.43}_{-0.25}$ & 1.21 (54) \\
b5 & 1138 & 2.08$^{+0.49}_{-0.33}$ & 0.44$^{+0.28}_{-0.19}$ & 1.70 (50) \\
b6 & 1478 & 2.23$^{+0.43}_{-0.31}$ & 0.37$^{+0.26}_{-0.17}$ & 1.37 (65) \\
b7 & 1250 & 1.99$^{+0.33}_{-0.25}$ & 0.29$^{+0.22}_{-0.13}$ & 1.13 (56) \\
b8 & 1385 & 2.30$^{+0.40}_{-0.26}$ & 0.63$^{+0.35}_{-0.25}$ & 1.25 (62) \\
b9 & 1056 & 1.81$^{+0.32}_{-0.21}$ & 0.27$^{+0.19}_{-0.13}$ & 1.00 (46) \\
b10 & 1562 & 2.10$^{+0.31}_{-0.28}$ & 0.38$^{+0.21}_{-0.15}$ & 0.94 (69) \\
b11 & 1181 & 1.96$^{+0.32}_{-0.25}$ & 0.38$^{+0.28}_{-0.16}$ & 1.20 (51) \\
b12 & 1101 & 1.48$^{+0.22}_{-0.09}$ & 0.23$^{+0.14}_{-0.09}$ & 0.89 (48) \\
c1 & 900 & 1.88$^{+0.34}_{-0.26}$ & 0.42$^{+0.25}_{-0.20}$ & 1.27 (39) \\
c2 & 2260 & 2.12$^{+0.20}_{-0.22}$ & 0.48$^{+0.20}_{-0.15}$ & 1.43 (97) \\
c3 & 5120 & 2.05$^{+0.14}_{-0.15}$ & 0.42$^{+0.11}_{-0.09}$ & 0.96 (204) \\
c4 & 1401 & 2.36$^{+0.46}_{-0.34}$ & 0.57$^{+0.38}_{-0.25}$ & 1.18 (62) \\
\hline
\end{tabular}
\caption{3C 66B: Best-fitting {\it mekal} models; the choice of regions is described in Section 3.1 and illustrated in Fig.~\ref{66bregs}. Errors are 1$\sigma$ for two interesting parameters.}
\label{66bspec}
\end{table*}

Fig.~\ref{66bimages} shows a contour plot of a smoothed image obtained by combining data from the three {\it XMM-Newton} cameras, as well as an adaptively smoothed combined image with radio contours overlaid. The morphology of the diffuse emission surrounding 3C 66B has a number of interesting features: there appear to be two deficits in surface brightness to the west and east of the centre, and there is a region of brighter emission at the edge of the eastern deficit. We tested the significance of the deficits by comparing the counts in these regions and in non-deficit regions at the same radius, and found the eastern deficit to be significant at the 4$\sigma$ level, and the western deficit to be significant at the 2$\sigma$ level. The overlaid radio contours on the adaptively smoothed image illustrate the relative positions of X-ray and radio features. There is a correspondence between the positions of the X-ray deficits and the radio lobes, and, in addition, there is a brighter blob of emission at the end of the eastern jet/lobe. The relationship between these radio and X-ray features is discussed in more detail in Section 4.1 below.

3C 66B has been observed in the X-ray with {\it Chandra} (Hardcastle, Birkinshaw \& Worrall 2001) and the observed X-ray emission was found to comprise an AGN component, emission from a jet, and extended emission attributed to hot gas extending to at least 40 arcsec from the core. The resolution of {\it XMM-Newton} means that the AGN and jet are not spatially separable from the extended emission in our data, and so we obtained a spectrum for a region consisting of the central 60 arcsec, to include all the AGN emission, for comparison with the {\it Chandra} results. A background spectrum was taken from a surrounding annulus to facilitate comparison (in contrast to the background method described above and used for all other spectra presented here). We fitted the data from the three {\it XMM-Newton} cameras with an absorbed power law plus {\it mekal} model, and found a best-fitting power-law energy index of $0.87^{+0.27}_{-0.17}$, consistent with the value of 1.14$\pm$0.05 of Hardcastle et al. (2001) , and best-fitting temperature for the extended component of 0.65$\pm0.05$ keV, also in agreement with their results (all errors quoted for our analysis are 1$\sigma$ confidence levels; those for {\it mekal} temperatures and abundances are 1$\sigma$ for two interesting parameters). The model used the Galactic $N_{H}$ value of 8.36 $\times 10^{20}$ cm$^{-2}$ (Murphy et al. 1996), gave a best-fitting abundance of 0.08$^{+0.36}_{-0.02}$ solar, and had a reduced $\chi^{2}$ of 0.98 (197 d.o.f.). We extracted a matching spectrum from the same region of the {\it Chandra} data, and found that a joint fit to all four spectra (MOS1, MOS2, pn, and {\it Chandra}) resulted in fits which did not differ significantly from the results quoted above. The high degree of consistency between the {\it XMM-Newton} and {\it Chandra} spectra suggests that there has not been significant variation of the AGN between the observations. The {\it Chandra} data better constrain the AGN spectral parameters because a single power-law model could be applied to the spatially separated spectrum. We measure a total unabsorbed flux (in the energy range 0.5 to 4.0 keV) from the power-law component (1.90$\pm0.47) \times 10^{-13}$ ergs cm$^{-2}$ s$^{-1}$, consistent with the total flux of core and jet components from Hardcastle et al.'s {\it Chandra} results of 2.31 $\times 10^{-13}$ ergs cm$^{-2}$ s$^{-1}$.

A spectrum was obtained for the entire extent of the diffuse X-ray emission surrounding 3C 66B out to a radius of 450 arcsec, excluding the central 60 arcsec. The best-fitting model has a temperature of 1.73$^{+0.03}_{-0.04}$ keV with an abundance of 0.28$\pm$0.03 ($\chi^{2}/n$ = 1.14 for 718 d.o.f.). The total X-ray luminosity (in the range 0.5 to 10 keV) was found to be (2.81$\pm$0.05) x 10$^{35}$ W. We applied the X-ray luminosity-temperature relation for groups described in Helsdon \& Ponman (2000) [found by Worrall \& Birkinshaw (2001) to be consistent with a sample of FR-I radio-galaxy environments], which gives an expected temperature of 0.79$\pm$0.03 keV, significantly lower than that obtained from our observations. We return to this point in Section 4.3.

\begin{figure*}
\begin{center}
\centerline{\hbox{
\epsfig{figure=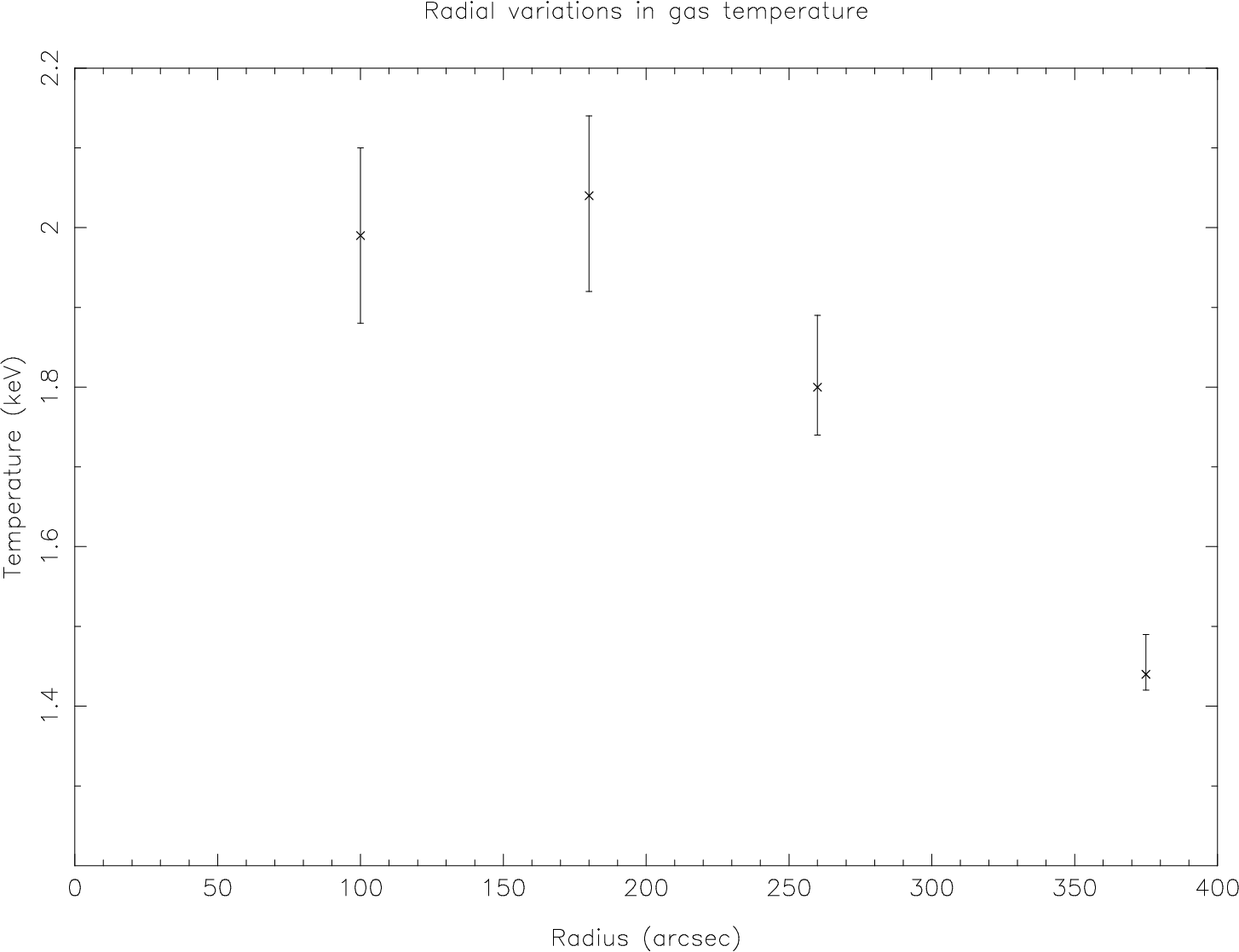,width=6.5cm,angle=270}
\epsfig{figure=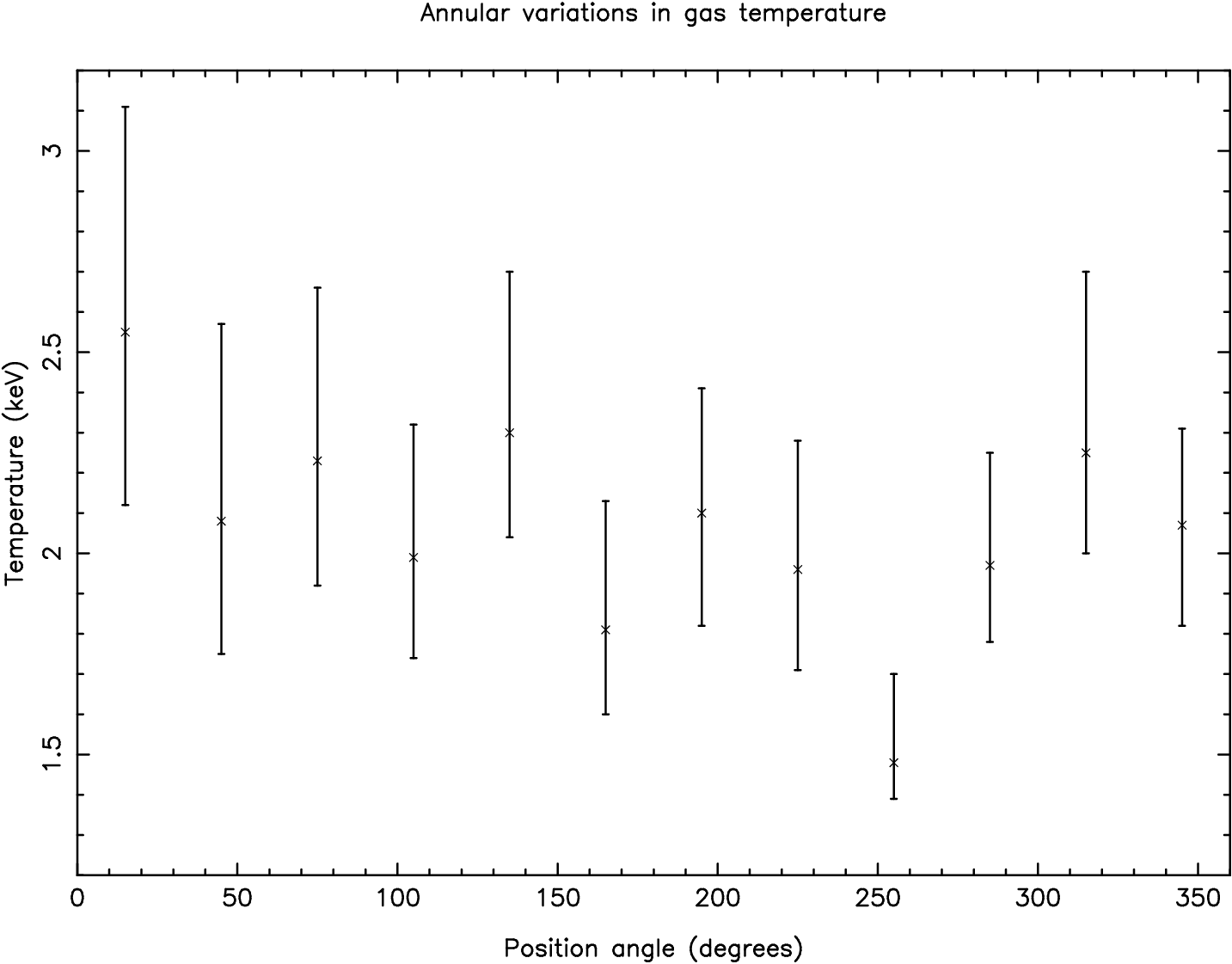,width=6.5cm,angle=270}}}
\caption{Plots of temperature variations in 3C 66B's environment as a function of radius and angle with abundace for the fits shown in Table~\ref{spec}.}
\label{66bplots}
\end{center}
\end{figure*}

\begin{figure*}
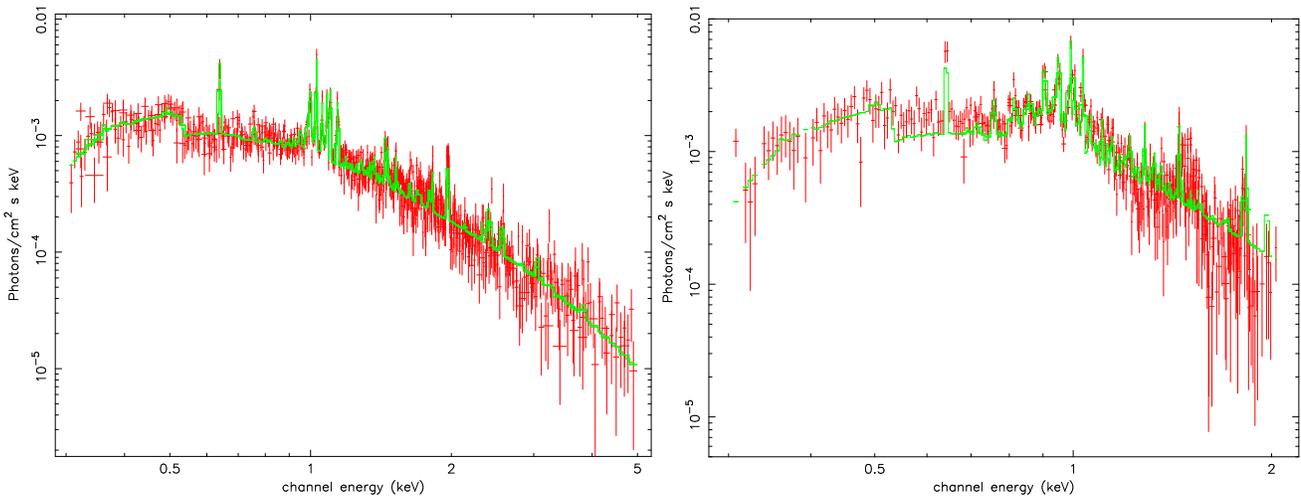

\begin{center}
\centerline{\hbox{
\epsfig{figure=66b_total.ps,width=6.5cm, angle=270}
\epsfig{figure=449_total.ps,width=6.5cm, angle=270}}}
\caption{Combined MOS1, MOS2 and pn spectra for extended emission from 3C 66B to a radius of 450 arcsec (l), and 3C 449 to a radius of 800 arcsec (r) with best-fitting {\it mekal} models of temperature 1.73 keV (3C 66B) and 0.98 keV (3C 449) as described in the text.}
\label{aspec}
\end{center}
\end{figure*}

As described in Section 2 above, a series of spectral extraction regions were chosen for more detailed analysis; these included annuli for radial temperature analysis, pie-slice regions for study of angular variations, and regions selected to study features of particular interest evident in the image of Fig.~\ref{66bimages}. The choice of extraction regions is illustrated in Fig.~\ref{66bregs}. The regions used for study of two other objects in the field of view, 3C 66A and another unidentified extended object (discussed in Sections 3.1.1 and 3.1.2), are also shown in this Figure.
The results of spectral analysis are given in Table~\ref{66bspec}, which shows the results of jointly fitting an absorbed {\it mekal} model to spectra from the MOS1, MOS2 and pn cameras in the energy range of 0.3 to 5.0 keV. The neutral hydrogen column density was fixed at 8.36 x 10$^{20}$ cm$^{-2}$ (Murphy et al. 1996). Temperature and abundance were allowed to vary. Best-fitting models are shown in Table~\ref{66bspec}, and plots of temperature as a function of angle and radius are shown in Fig.~\ref{66bplots}. Fig.~\ref{aspec} shows the best-fitting {\it mekal} model for the total emission.

\begin{figure*}
\begin{center}
\centerline{\hbox{
\epsfig{figure=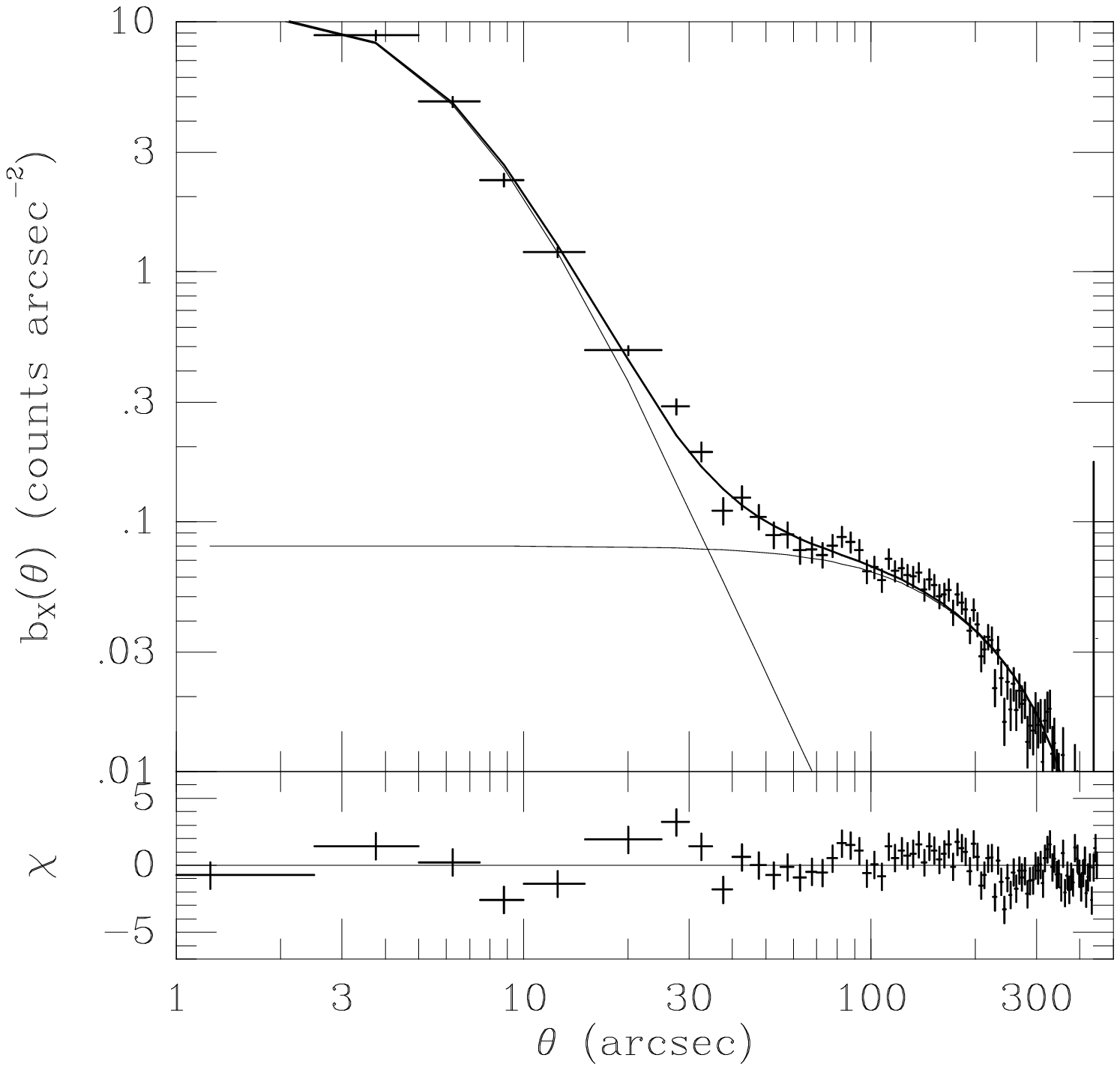,width=8.0cm}
\epsfig{figure=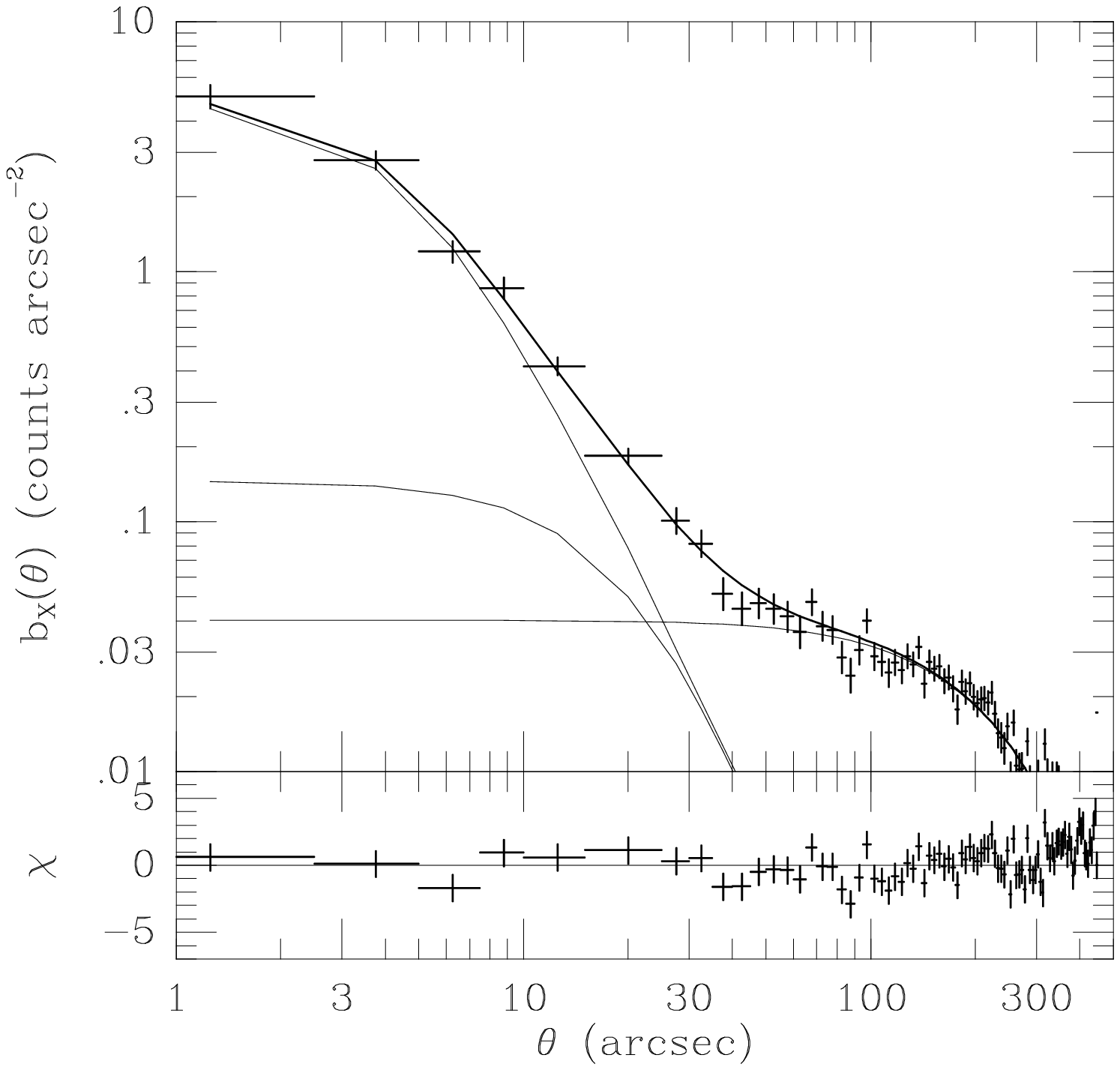,width=8.0cm}}}
\caption{a) Radial surface brightness profile for 3C 66B using pn data only; model is the best-fitting point-source plus $\beta$-model to the MOS1, MOS2 and pn profiles, with $\beta$ = 0.375 and r$_{c}$ = 179.6 arcsec, b) MOS1 profile with the same model as for a), but with a second $\beta$-model component having $\beta$ = 0.9 and r$_{c}$ = 20 arcsec}
\label{66brad}
\end{center}
\end{figure*}

Radial surface brightness profiles were extracted for each camera and fitted with convolved point-source plus $\beta$-models as described in Section 2 above. The $\chi^{2}$ values for the three cameras were combined for each fit to determine the best joint fit to all the three profiles. We found a best-fitting model with $\beta$ = 0.375$^{+0.125}_{-0.75}$, core radius = 179.6$^{+50}_{-40}$ arcsec with $\chi^{2}/n$ of 1.88 (267 d.o.f.) (errors are 1$\sigma$ for two interesting parameters). An unacceptable fit is not surprising, as the images and spectral analysis show that there is clearly small-scale structure in the extended emission. Fig.~\ref{66brad} shows the pn radial profile with the best-fitting model to the individual profiles. There are systematically positive residuals in the fits at small radii, which could be due to a galaxy-scale atmosphere. Our spectral results, and the {\it Chandra} data of Hardcastle et al. (2001) indicate that a galaxy-scale atmosphere would be expected in addition to the larger group-scale gas fitted by the model described above. We fitted a second $\beta$-model component by holding the large $\beta$-model fixed, and found that including a model with $\beta$ = 0.9 and core radius of 20 arcsec significantly improved the fit for both MOS1 and MOS2 (though the fit is still unacceptable for the reasons explained above), and resulted in a slight, but not significant improvement for the pn fit. The two-$\beta$-model fit for MOS1 is shown in Fig.~\ref{66brad}. The presence of this smaller-scale component is consistent with the {\it Chandra} data; however, its parameters cannot be well-constrained by our {\it XMM-Newton} data, where much of the emission on these scales is hidden by the PSF, or by the {\it Chandra} data, since its surface brightness is rather low.

\subsubsection{3C 66A}

3C 66A is a BL Lac object at a redshift of 0.44, seen 6.5 arcmin to the north-west of 3C 66B in the images of Fig.~\ref{66bimages}. Initially thought to be related to 3C 66B, it was soon determined to be a separate object (Mackay et al. 1971; Northover 1973), and was optically identified by Wills \& Wills (1974). Its spectrum was extracted for the three cameras, and an absorbed power law model was fitted in the energy range of 0.3 to 7.0 keV. Allowing the Galactic $N_{H}$ to vary, a photon index of $2.47\pm0.03$ was obtained for a value of $N_{H}$ = (6.9$\pm0.5$) $\times$ 10$^{20}$ cm$^{-2}$ ($\chi^{2}/n=1.17$ for 668 d.o.f.). This value for $N_{H}$ is slightly lower than the value of 8.36 x 10$^{20}$ cm$^{-2}$ used in the 3C 66B spectral fits above (Murphy et al. 1996). Fixing the $N_{H}$ at the Galactic value, a photon index of 2.55$\pm$0.01 was obtained ($\chi^2/n=1.19$ for 669 d.o.f.). The photon indices for both fits are consistent with the results of Sambruna et al. (1994), Fossati et al. (1998), and Worrall \& Wilkes (1990) [as tabulated by Donato et al. (2001)]. The main residual component in these fits was at high energies; including a flat or heavily absorbed second power-law component significantly improves the fit. For a second power law with photon index of -0.3$\pm$0.5, we obtain $\chi^{2}/n$ of 1.11 (667 d.o.f.); a heavily absorbed power law (N$_{H}$ = 22$^{+14}_{-11} \times 10^{22}$ cm$^{-2}$) with energy index 2.1$^{+2.6}_{-1.3}$ gives $\chi^{2}/n$ = 1.10 (666 d.o.f). The photon index and normalization of the single component fit with fixed Galactic column density were adopted, and this model led to a 1-keV flux density of 0.378 $\mu$Jy, which is lower than the values obtained in the three earlier studies by a factor of 3 to 5. This amount of variability is not implausible, as optical monitoring shows variability of 1.5-2 mag (e.g. Takalo et al. 1996). There was no significant variability evident in the light-curve of 3C 66A over the course of our observation. We fitted a radial surface-brightness profile to the emission from 3C 66A, and this was adequately fit by a point-source model with no extended emission. Although this is a bright object, the count rates are too low for pile-up to occur, as is confirmed by {\sc sas} task {\it epatplot}, so we conclude that the lower X-ray flux density is a real effect of source variability.

\subsubsection{Nearby cluster}

The images of Fig.~\ref{66bimages} show an extended object $\sim$ 7 arcmin to the south-east of 3C 66B. We find 4874 background-subtracted counts (MOS1, MOS2 + pn) in a 78 arcsec circle. The position of this emission does not coincide with any previously detected cluster. An absorbed mekal model was fitted in the energy range 0.3 to 7.0 keV, with an abundance of 0.4 solar, allowing redshift and temperature to vary. The best fit temperature was 3.7$^{+0.5}_{-0.3}$ keV, at a redshift of $0.35^{+0.01}_{-0.04}$; $\chi^{2}/n$ for this fit was 1.33 for 174 d.o.f. The unabsorbed X-ray luminosity (between 2 and 10 keV) was found to be (1.66$^{+0.02}_{-0.09}) \times 10^{37}$ W. Applying the X-ray luminosity-temperature relation (Arnaud \& Evrard 1999) results in an expected temperature of 3.0 keV which is in approximate agreement with our best fit value. The X-ray luminosity/brightest cluster member relation of Edge \& Stewart (1991) predicts that the brightest cluster galaxy should have a visual magnitude of $\sim$ 19. DSS images of this field show a faint extended object at the position of the centre of emission, with several other objects of similar magnitude nearby. Therefore we conclude that this is a strong candidate for a new X-ray-detected cluster of galaxies at $z = 0.35$ at $\alpha$ = 02:23:34, $\delta$ = +42:54.

\subsection{3C 449}

\begin{table*}
\centering
\begin{tabular}{@{}l|llll@{}}
\hline
Region  & Net counts & Temp (keV) & Abundance & $\chi^{2}/n$ (d.o.f.) \\
\hline
tot & 51276 & 0.98$\pm$0.02 & 0.13$^{+0.01}_{-0.02}$ \\ 
a1 & 9665 & 1.39$\pm$0.02 & 0.33$^{+0.04}_{-0.03}$ & 1.21 (331) \\
a2 & 8256 & 1.23$^{+0.05}_{-0.06}$ & 0.15$\pm$0.03 & 1.27 (287) \\
a3 & 8076 & 1.11$\pm$0.03 & 0.13$\pm$0.02 & 1.34 (278) \\
a4 & 9372 & 1.04$^{+0.03}_{-0.04}$ & 0.12$\pm$0.02 & 1.13 (294) \\
b1 & 3156 & 1.33$^{+0.06}_{-0.08}$ & 0.17$^{+0.06}_{-0.04}$ & 0.95 (128) \\
b2 & 2717 & 1.12$^{+0.06}_{-0.04}$ & 0.19$^{+0.05}_{-0.04}$ & 1.16 (111) \\
b3 & 2396 & 1.07$\pm$0.04 & 0.17$^{+0.05}_{-0.04}$ & 0.93 (96) \\
b4 & 3139 & 1.11$^{+0.05}_{-0.03}$ & 0.23$^{+0.06}_{-0.04}$ & 1.20 (124) \\
b5 & 2951 & 1.24$^{+0.07}_{-0.09}$ & 0.29$^{+0.09}_{-0.08}$ & 1.34 (117) \\
b6 & 2728 & 1.31$^{+0.07}_{-0.10}$ & 0.17$^{+0.07}_{-0.05}$ & 1.21 (111) \\
b7 & 2425 & 1.12$^{+0.09}_{-0.08}$ & 0.08$\pm$0.03 & 1.08 (102) \\
b8 & 2304 & 1.12$^{+0.15}_{-0.07}$ & 0.09$^{+0.05}_{-0.02}$ & 1.27 (93) \\
b9 & 2496 & 1.11$^{+0.10}_{-0.07}$ & 0.06$^{+0.03}_{-0.02}$ & 1.25 (101) \\
b10 & 2504 & 1.26$^{+0.09}_{-0.11}$ & 0.17$^{+0.07}_{-0.06}$ & 1.10 (103) \\
b11 & 3756 & 1.22$^{+0.08}_{-0.10}$ & 0.12$^{+0.04}_{-0.03}$ & 1.12 (149) \\
b12 & 3896 & 1.29$\pm$0.07 & 0.18$^{+0.06}_{-0.04}$ & 0.96 (153) \\
c1 & 13240 & 1.01$\pm$0.03 & 0.12$\pm$0.02 & 1.24 (351) \\
\hline
\end{tabular}
\caption{3C 449: Best-fitting mekal models; the choice of regions is described in Section 3.2 and illustrated in Fig.~\ref{449regs}.}
\label{449spec}
\end{table*}

Fig.~\ref{449images} shows a contour plot of a smoothed image obtained by combining data from the three {\it XMM-Newton} cameras, as well as an adaptively smoothed combined image with radio contours overlaid. The radio-X-ray overlay illustrates the presence of a deficit at the position of the southern radio lobe; we tested the significance of this deficit by comparing the count rate with that of a region at similar radius in the direction perpendicular to the jet axis, and found it to be significant at greater than 5$\sigma$.

Hardcastle et al. (1998) presented {\it ROSAT} imaging and spectral analysis of the gas surrounding 3C 449. We took a spectrum of the full extent of the X-ray emission out to a radius of 420 arcsec to compare with the {\it ROSAT} results. We found a best-fitting {\it mekal} model with temperature of 0.98$\pm$0.02 keV, consistent with their results at the 2$\sigma$ level, and a luminosity in the energy range 0.2 - 1.9 keV of (1.73$\pm0.05) \times 10^{35}$ W, consistent with their value of 2.1 $\times$ 10$^{35}$ W (corrected for the different cosmology we use here). In the energy range 0.5 to 10 keV, we measure a luminosity of (1.98$\pm0.06) \times 10^{35}$ W. The luminosity-temperature relation of Helsdon \& Ponman (2000) predicts a temperature of 0.73$\pm$0.05 keV, lower than the value we measure. We took a spectrum of the central 60 arcsec to look for an AGN component; however, the best-fitting {\it mekal} plus power-law model has a photon index of 0.63$^{+0.53}_{-0.92}$ (with a temperature of 1.37$^{+0.26}_{-0.19}$ and $\chi^{2}/n$ of 1.26 for 93 d.o.f.). If the photon index is held fixed at the more physically plausible value of 2 (i.e. an energy index of 1), a $\chi^{2}/n$ of 1.27 (for 94 d.o.f.) is obtained. This is not a significantly worse fit; therefore we cannot constrain usefully the power-law index with these data. Taking the second model with fixed power-law index, we find a flux from AGN and jet components of 7.7 $\times 10^{-14}$ ergs cm$^{-2}$ s$^{-1}$, a factor of 3 lower than that of 3C 66B. This is consistent with the flux in the point-source component of our radial profiles described below.

\begin{figure*}
\begin{center}
\centerline{\hbox{
\epsfig{figure=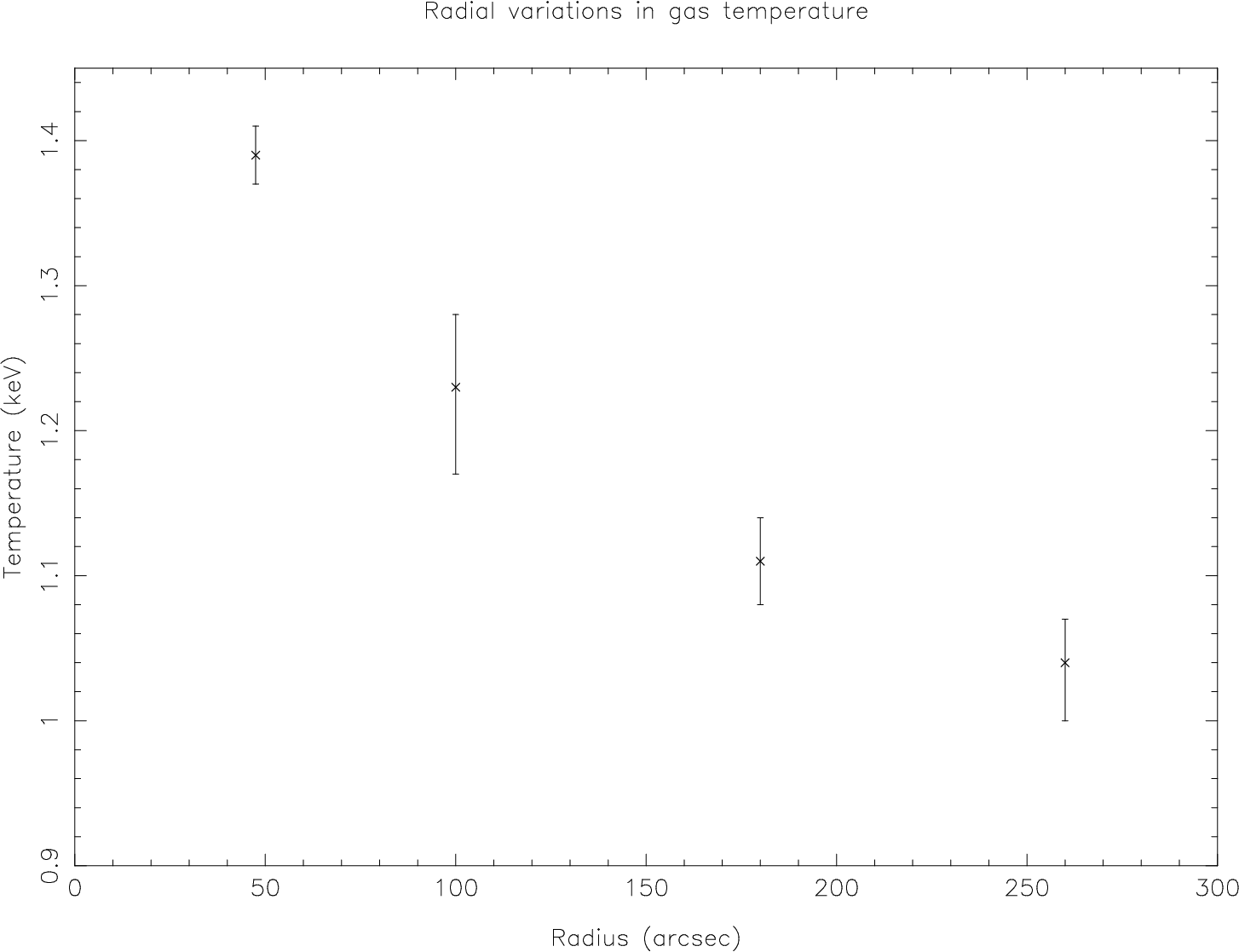,width=6.5cm,angle=270}
\epsfig{figure=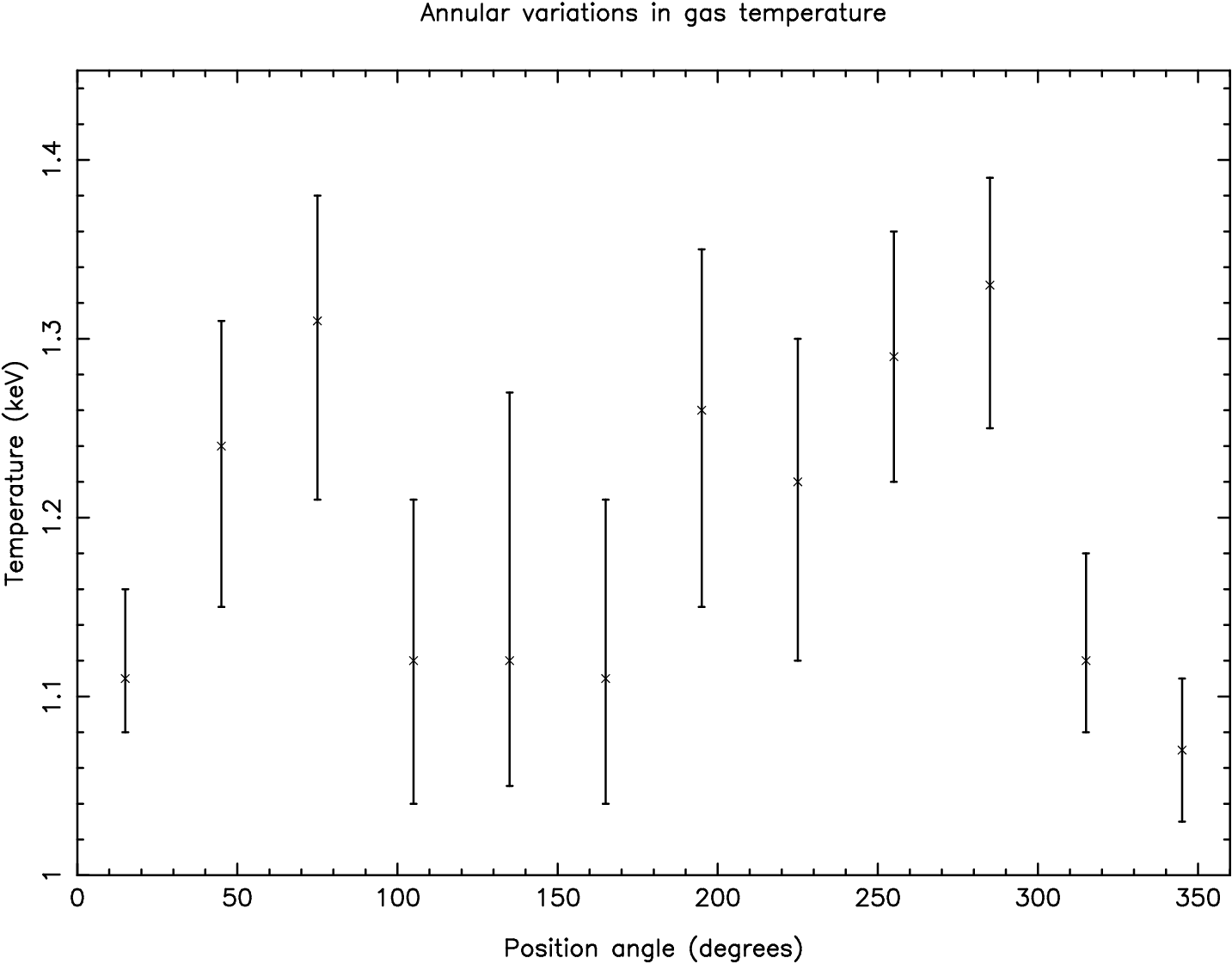,width=6.5cm,angle=270}}}
\caption{Plots of temperature variations in 3C 449's environment as a function of radius and angle for the fits shown in Table~\ref{449spec}.}
\label{449plots}
\end{center}
\end{figure*}

We defined a set of similar spectral extraction regions to those used for 3C 66B to look for radial and angular temperature structure. The choice of regions is illustrated in Fig.~\ref{449regs}. Fits were performed in the same way as for 3C 66B, using a fixed Galactic column density of 11.8 $\times$ 10$^{20}$ cm$^{-2}$ (Murphy et al. 1996): we found that varying the column density did not significantly improve our fits. Table~\ref{449spec} shows the best-fitting {\it mekal} model fits for fixed and free abundance. Plots of temperature variations, shown in Fig.~\ref{449plots} illustrate that the temperature is lower at position angles of around 180 and 360 degrees, corresponding to the position of the radio jets and lobes, whereas higher temperatures are found where there is brighter X-ray gas, perpendicular to the radio axis. The temperature is also found to decrease with radius, from 1.39 keV at a radius of 45 arcsec to 1.04 keV at a radius of 260 arcsec.

\begin{figure}
\begin{center}
\epsfig{figure=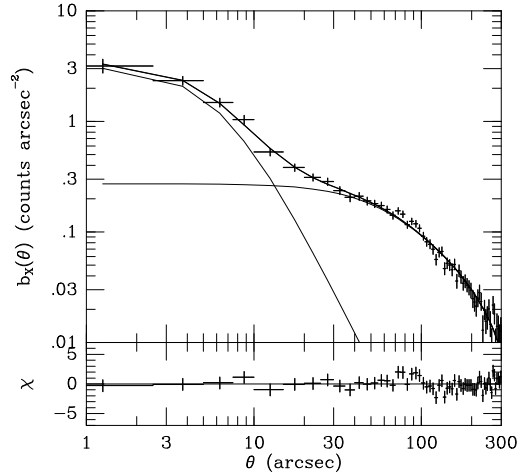,width=8.0cm}
\caption{Radial surface brightness profile for 3C 449 using pn data only; the model is the best-fitting point-source plus $\beta$-model to the MOS1, MOS2 and pn profiles, with $\beta$ = 0.375 and r$_{c}$ = 48.4 arcsec}
\label{449rad}
\end{center}
\end{figure}

Radial surface brightness profiles were extracted for each camera and fitted with convolved point-source plus $\beta$-models as for 3C 66B. We found a joint best-fitting model with $\beta$ = 0.375$\pm$0.025, core radius = 48.4$^{+7.6}_{-10.5}$ arcsec with $\chi^{2}/n$ of 1.27 (267 d.o.f.). As in 3C 66B, the unacceptable fit is likely to be due to the clumpiness of the environment. Fig.~\ref{449rad} shows the pn radial profile with the combined best-fitting model. The radial profiles show no evidence for a galaxy-scale atmosphere.

\section{Discussion}

\subsection{Interactions of radio plasma and X-ray-emitting gas}

\subsubsection{3C 66B}

The contour plot and adaptively smoothed image of 3C 66B (Fig.~\ref{66bimages}) show strong evidence for interactions between a low-power (FR-I) radio galaxy and its environment. The presence of deficits in the X-ray surface brightness coincident with the extended radio structure provides evidence both for the influence of the X-ray gas on the lobes, and for the displacement of X-ray-emitting gas. Hot gas appears to surround the eastern radio lobe, which has a sharp edge and `wall' at the end of the jet; in contrast, the more diffuse western lobe is relatively free of gas. This suggests that the eastern lobe has been constrained into its present shape by the surrounding hot gas, whereas the absence of gas at similar distances from the core on the opposite side of the source left the western jet relatively free to propagate. That the eastern lobe is the shorter of the two by a factor of $\sim$2.5 is consistent with the suggestion that the environment has slowed the lobe expansion. Emission round the edges of the western lobe for a short distance indicates that the expanding lobe has pushed material out of the way; however, the lobe morphology suggests that the gas was originally less dense than on the eastern side, allowing the western lobe to expand faster than its counterpart.

A particularly interesting feature in 3C 66B's X-ray-emitting environment is the bright blob of gas at the end of the eastern jet, just beyond the sharp boundary in the radio emission. As shown in Fig.~\ref{66bimages}, the radio jet appears to end suddenly, and lobe emission appears below and to the south of the blob of gas. However, it is unclear whether the blob of gas has been pushed out from near the centre of the radio galaxy, or whether the lobe has only recently encountered a denser clump of gas, which we find to have a mass of 4 $\times$ 10$^{9}$ M$_{\sun}$ (obtained from the volume emission measure of the best-fitting spectrum).

Without detailed hydrodynamical simulations it is difficult to determine the dynamics of this gas blob in the atmosphere. If it originated close to the centre of 3C 66B, we can determine a lower limit to the time required for the jet to push the blob out to its current distance from near the centre by neglecting the effect of gravity, and assuming that all of the underlying jet's thrust acts on the blob. Assuming a jet power of 3.5 $\times 10^{37}$ W [scaling the jet power of 3C 31 from models of Laing \& Bridle (2002) by the ratio of low-frequency radio luminosity in the two objects], we find that it would take 1.4 $\times 10^{8}$ years. This is likely to be less than the age of the source, as discussed in Section 4.3. However, gravity is likely to significantly increase the time required. 

We calculated the mass swept out by the front of the expanding radio lobe to be $\sim$4 $\times$ 10$^{9}$ M$_{\sun}$, the same as the mass we calculate for the blob. This mass was determined by integrating the density of the $\beta$-model atmosphere, assuming the volume swept out by the lobe is a cone of half-opening angle 45$^{\circ}$. We also find that the temperature increase produced by adiabatic compression of the gas evacuated from the lobe to the current blob volume is consistent at the 2$\sigma$ level with our measured temperature of 2.36$^{+0.46}_{-0.34}$ keV (significantly hotter than the surrounding medium) assuming $\gamma = 5/3$. However, it seems unlikely that this could explain its origin, as the gas would be expected to spread out, rather than form such a dense compact blob. 

The position of this blob and the radio-source morphology at the end of the lobe are suggestive of a scenario where the gas forms an obstacle choking off the flow and forcing the lobe to expand laterally. Either the jet could have encountered a denser clump of gas which is part of the environment, or an infalling clump of gas. There is no evidence that the blob has a different redshift than the surrounding material. Both scenarios are physically plausible and consistent with the radio-source morphology of 3C 66B. 

\subsubsection{3C 449}

Evidence was presented in Hardcastle et al. (1998) for a deficit in X-ray surface brightness coincident with the southern radio lobe of 3C 449 and Fig.~\ref{449images} confirms the presence of this deficit. The narrow northern lobe is free of X-ray-emitting gas at 100 kpc from the core, whereas there is more gas to the sides and below the more rounded southern lobe (unfortunately the southern lobe extends to the edge of the field of view, so the amount of gas below the radio lobe is not clear). In both cases there are deficits in X-ray surface brightness at the lobe positions. The 1.4-GHz radio map of Feretti et al. (1999) shows the southern lobe to have more internal structure than appears in our lower resolution map. However, the rounded appearance of this lobe suggests a similar scenario to that described above for 3C 66B: the southern lobe expansion along the jet axis has been slowed by the denser material we see producing the higher level of X-ray emission in that direction.  Therefore, our images of the X-ray-emitting gas surrounding 3C 449 give further support for the idea that the X-ray environment plays an important role in determining the morphology of the radio lobes of low-power radio galaxies. The radio maps of Feretti et al. (1999) also show smaller ``bulges'' in both jets nearer to the core, which could be due to the jets crossing the boundary between a denser galaxy atmosphere and the group gas. We find that the gas in the surrounding rim of 3C 449's southern lobe has a mass of 1.1 $\times 10^{10}$ M$_{\sun}$ (we took the spectrum from a section of the rim using local background subtraction, then extrapolated this mass to a sphere of thickness 34 arcsec; this is likely to be an overestimate, since there will be no rim on the core side of the lobe). As shown in Table~\ref{449spec}, the rim has a temperature of 1.01$\pm$0.03 keV, consistent with the temperature in the outer regions of the extended gas. We calculate that the amount of gas swept up by the lobe front is $\sim 6 \times 10^{10}$ M$_{\sun}$. It is therefore plausible that most of the rim gas has been evacuated from the lobe and moved to its current position by lobe expansion.

\subsection{Pressure balance between radio lobes and X-ray environment}

\begin{figure*}
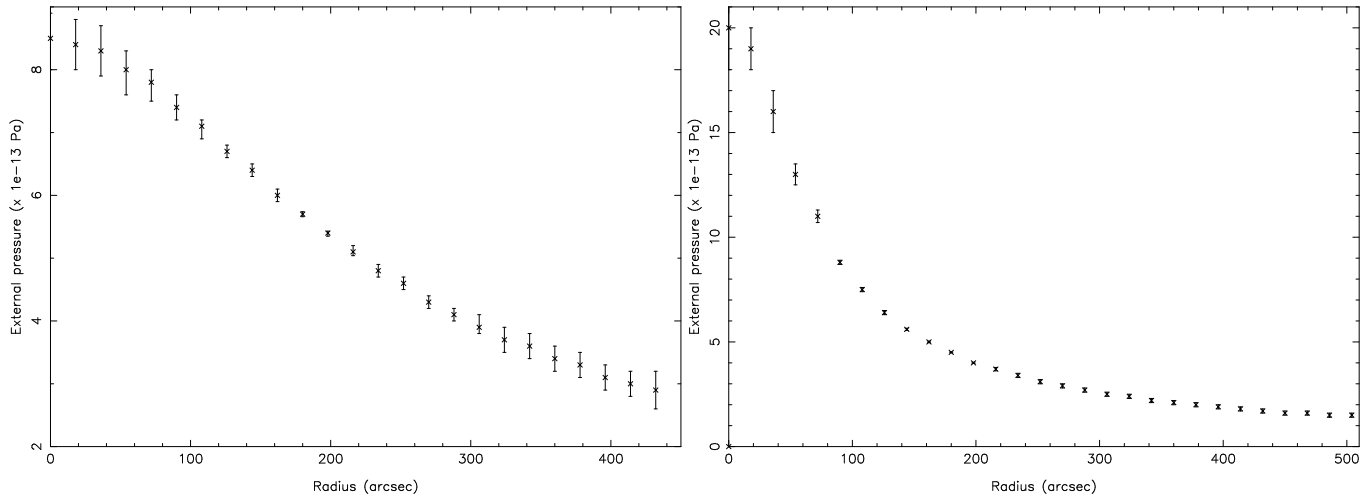

\begin{center}
\centering{\hbox{
\epsfig{figure=66b_press.ps, width=6.5cm, angle=270}
\epsfig{figure=449_press.ps, width=6.5cm, angle=270}}}
\caption{Plots of external pressure in the X-ray-emitting gas around 3C 66B (left) and 3C 449 (right) as a function of distance from the centre. Errors are 1$\sigma$, determined from errors in the best-fitting $\beta$-model parameters, which assumes an isothermal environment of 1.73 keV (3C 66B) and 1.15 keV (3C 449).}
\label{press}
\end{center}
\end{figure*}

The radio lobes of low-power radio galaxies are thought to be expanding subsonically on the largest scales, so that the internal pressure within the radio lobes should be similar to, but slightly higher than that of the external environment. However, internal radio-lobe pressures obtained by assuming an electron filling factor of unity, equipartition of energy in particles and magnetic fields, and that the only contribution to pressure comes from the population of synchrotron-emitting electrons, are found to be significantly {\it lower} than the external pressures obtained from X-ray measurements (e.g. Morganti et al. 1988; Worrall et al. 1995; Worrall \& Birkinshaw 2000). If the lobes are expanding supersonically, these problems are exacerbated (more detailed discussion of expansion speeds follows in Section 4.3). We determined radial pressure profiles in the X-ray environments of 3C 66B and 3C 449; these are shown in Fig.~\ref{press}. Internal radio-lobe pressures were determined using the code of Hardcastle et al. (1998), choosing an electron energy spectrum with a power-law number index of 2, minimum energy of 5 $\times 10^{6}$ eV and maximum energy of 6 $\times 10^{11}$ eV.

For 3C 66B, we find the external pressure acting on the eastern lobe of 3C 66B at a radius of 75 arcsec to be a factor of $\sim$24 higher than the equipartition internal lobe pressure, and for the southern lobe of 3C 449 we find the external pressure at a distance of 350 arcsec to be $\sim$16 above the internal lobe pressure. The result for 3C 449 agrees with Hardcastle et al.'s {\it ROSAT} results, and both results are consistent with {\it ROSAT} measurements for similar sources (e.g. Worrall et al. 1995). Therefore, as the radio lobes cannot be underpressured, we must consider which of the assumptions used to determine the internal lobe pressures are incorrect. 

\begin{figure}
\begin{center}
\epsfig{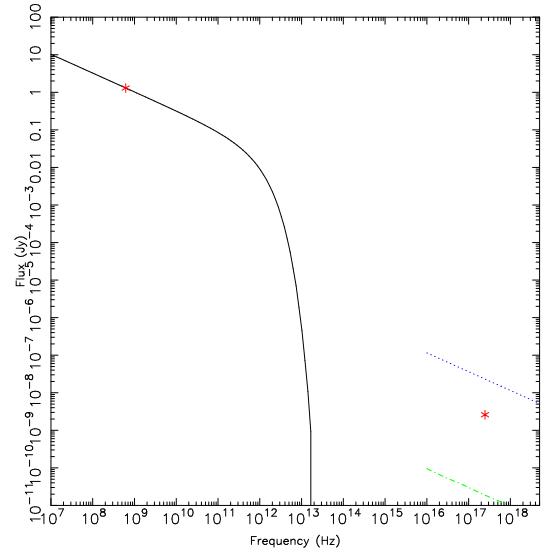}
\caption{Plot of the modelled spectrum of 3C 449's southern lobe for magnetic field strength of 20 pT (to achieve pressure balance with external environment). The solid line is the synchrotron emission, passing through a radio data point. The second data point is an upper limit to X-ray inverse-Compton emission as described in the text, which is significantly below the dashed line indicating the expected inverse Compton emission from CMB photons scattered by the population of synchrotron-emitting eletrons. The lower dashed line is the expected level of synchrotron self-Compton emission}
\label{magdom}
\end{center}
\end{figure}

In order for the population of synchrotron-emitting electrons to provide the necessary pressure without a significant contribution from other particles, conditions in the radio lobes of the two objects must deviate substantially from equipartition.  For 3C 66B, we find an equipartition magnetic field strength of 0.34 nT (1 nT = 10$\mu$G). To obtain pressure balance, field strengths of 30 pT or 2.0 nT are required. In 3C 449 we obtain similar results: the equipartition magnetic field is 0.23 nT, whereas field strengths of 20 pT or 1.2 nT are required for pressure balance. However, from spectral fitting to the X-ray deficit regions coincident with these radio lobes we obtain an upper limit on the flux from inverse Compton emission of 10 nJy (3C 66B) and 3 nJy (3C 449), which leads to lower limits on magnetic field strength of 90 pT (3C 66B) and 80 pT (3C 449). With such fields, the lobes would still be underpressured by factors of 5 (3C 66B) and 7 (3C 449). Fig.~\ref{magdom} shows that the level of inverse Compton emission expected from the lobes of 3C 449 for a field strength of 20 pT (required for pressure balance in the particle-dominated regime) is significantly higher than the upper limit of a power-law component of X-ray flux, as taken from fitting a {\it mekal} plus power-law model to the spectrum from the lobe region. Therefore we conclude that in the scenario where no other particles are present, if the lobes of these two objects deviate from equipartition conditions, it must be in the direction of magnetic domination, requiring field strengths of $\sim$ 1-2 nT. Such magnetic domination is not found where direct measurements of magnetic field strengths have been made from X-ray inverse Compton emission. In no case is the field strength significantly higher than the equipartition value; Feigelson et al. (1995) found a level of X-ray inverse Compton in the lobes of Fornax A slightly higher than, but consistent with equipartition, and Tashiro et al. (1995) found the lobes of PKS 1343-601 to be particle dominated. 

Our results also constrain pressure contributions from non-radiating particles. Although the presence of some entrained material in the lobes is necessary (e.g. from models of jet deceleration), the X-ray deficits rule out a significant contribution to pressure from a thermal gas component at the temperature of the surrounding atmosphere. The presence of sufficient cold gas would be expected to produce significant Faraday depolarization of the lobes, which is not observed (e.g. Feretti et al. 1999), although it is possible to devise magnetic field structures where such material could be present but does not result in significant depolarization (Laing 1984). However, it is implausible that large quantities of material could be hidden in this way. Entrained material could be heated (by the entrainment process, or by energy transfer from interactions with the relativistic electrons), and so could provide sufficient pressure without producing detectable X-ray emission: e.g. in 3C 66B, the missing pressure might be provided by gas at $\sim$5 keV. Missing pressure could also be hidden in an extra population of low-energy electrons, producing inverse Compton emission at a level we could not detect. However, these electrons must have $\gamma \sim 1-200$, requiring the power-law number index to steepen to 3 below $\sim$ 10$^{8}$ eV, then to drop off sharply at energies of $\sim 5 \times 10^{5}$ eV, since large populations of electrons at these energies would also be expected to result in depolarization. This model is somewhat implausible. There could also be a contribution from relativistic protons, although this would require large proton to electron ratios of $\kappa \sim$260 for 3C 66B, and $\sim$130 for 3C 449. It also possible that there is a plasma filling factor $\phi$ less than unity; however, to achieve pressure balance, $\phi \sim$ 1/260 (3C 66B) and 1/130 (3C 449) would be required. More detailed discussion of this scenario is found in Hardcastle et al. (2000). 

We have assumed that the source is in the plane of the sky, which may not be the case. As described in Hardcastle et al. (1998), if the radio source is at an angle of $\theta$ to the line of sight, the minimum internal pressure falls as $(\sin\theta)^{4/7}$, so that small angles to the line of sight are required to achieve pressure balance. In 3C 66B, an angle of 7$^{\circ}$ to the line of sight is required, and in 3C 449 an angle of 11$^{\circ}$ is required. In both cases these angles are implausible because they imply extremely large physical sizes. Additionally, Feretti et al. (1999) find that in 3C 449, $\theta>75^{\circ}$, which argues strongly against such an explanation.

\subsection{Radio-source heating of the X-ray environment}

In Section 3.1 we showed that the environments of 3C 449 and particularly 3C 66B are significantly hotter than predicted temperatures from the luminosity-temperature relation of Helsdon \& Ponman (2000). It is interesting to speculate on possible explanations for these increased temperatures. Expanding radio lobes would be expected to put a significant proportion of their energy budget into $P$d$V$ work on their environment. Only one example of radio-source heating of the environment has been directly detected so far: shock-heating of the environment of Cen A (Kraft et al. 2003). In the case of 3C 66B, determining the atmosphere's heat capacity using $\Delta E/\Delta T = (3/2)nk$, where $n = 2.18 n_{p}$, which is the total number of particles, where $n_{p}$ is the number of protons (determined from the volume emission measure obtained from spectral fitting to the entire environment to a radius of 450 arcsec), we calculate that the energy needed to heat the entire environment by 0.9 keV (from the predicted temperature of $\sim$0.8 keV to the observed 1.7 keV) is $\sim 2.4 \times 10^{53}$ J. If we assume that the radio jets of 3C 66B have a luminosity of $3.5 \times 10^{37}$ W [scaling the jet luminosity of 3C 31 (Laing \& Bridle 2002) by the ratio of low-frequency radio luminosities of the two objects] and that 1/3 of this energy goes into heating the environment by $P$d$V$ work (since pressure is 1/3 $\times$ energy density), we find that it would take $\sim 3 \times 10^{8}$ years for 3C 66B to heat its environment by this amount. Using 1.4- and 8-GHz radio data, we fitted a Jaffe \& Perola (1973) aged synchrotron spectrum with injection index of 0.5, in a magnetic field of 0.3 nT, to material at the far edge of the eastern lobe, we obtain a spectral age for 3C 66B of $\sim 10^{8}$ years. In order to provide the necessary energy input for a temperature increase of 0.9 keV in 10$^{8}$ years, the jets would need to be $\sim$ 3 times more powerful than the value we use above (or a larger proportion of the jet energy would have to go into radio-source heating). 

However, spectral ages are based on the spectrum of the oldest measured radio material in the lobes. Such ages should be considered as lower limits, as there may be even older material which is no longer emitting. To investigate whether 3C 66B and 3C 449 could be older than their measured spectral ages, we used the sizes of the two radio sources to estimate their ages, assuming they are in the plane of the sky (thus providing a lower limit to size and hence age). The distance from the radio core to the western lobe edge of 3C 66B (in Fig.~\ref{66bimages}) is 175 kpc, and the sound speed in its environment is $\sim7 \times 10^{5}$ m s$^{-1}$, which means that if the lobe front is moving subsonically, the source age must be at least 3 $\times 10^{8}$ years. In the case of 3C 449, the northern edge (in Fig.~\ref{449images}) is 290 kpc from the core and the sound speed is $\sim6 \times 10^{5}$ m s$^{-1}$, which implies a source age exceeding 5 $\times 10^{8}$ years. Interestingly, in both cases these values are close to the times required to raise the group temperatures from the predicted radio-quiet temperatures to those observed. There are two reasons to favour radio-source heating as an explanation of hotter group gas: firstly, in both sources the morphology suggests that the lobes are unlikely to be supersonic; secondly, were the lobes supersonic, they would be expected to be overpressured, increasing the need for additional pressure components or magnetic field as described in Section 4.2. Therefore we suggest that FR-I sources are likely to be older than typical spectral ages suggested by earlier work (e.g. Andernach et al. 1992; Parma et al. 2002). This means that both 3C 66B and 3C 449 could be old enough to have heated their environments by the amount necessary to produce the observed temperature increases. Eilek (1996) reached similar conclusions about radio-source ages from a more detailed dynamical analysis.

The luminosity-temperature relationship we have used is based on an extrapolation of the relation for clusters to radio-galaxy environments (Worrall \& Birkinshaw 2000). If radio-source heating is occurring in the environment of 3C 66B, we might expect that it could affect the temperatures of other environments in their sample, so that we are not comparing 3C 66B with unheated environments. The larger X-ray group sample of Helsdon \& Ponman (2000) also contains a significant proportion of radio-galaxy host groups. They find a relation for groups which differs from the cluster relation. Therefore, as the luminosity-temperature relation for groups which do not contain radio sources is unknown, we decided to investigate this by extracting radio-loud and radio-quiet subsamples from the group sample of Helsdon \& Ponman. The criteria used in selecting the radio-quiet subsample was that there were no radio sources associated with the group within 2 core radii of the central galaxy whose 1.4-GHz luminosity was greater than 10 per cent of the value we obtained for NGC 3665, a radio galaxy (B2 1122+39) in the Helsdon \& Ponman sample. We used NED, Simbad, FIRST and NVSS to check for radio sources. 

\begin{figure}
\begin{center}
\epsfig{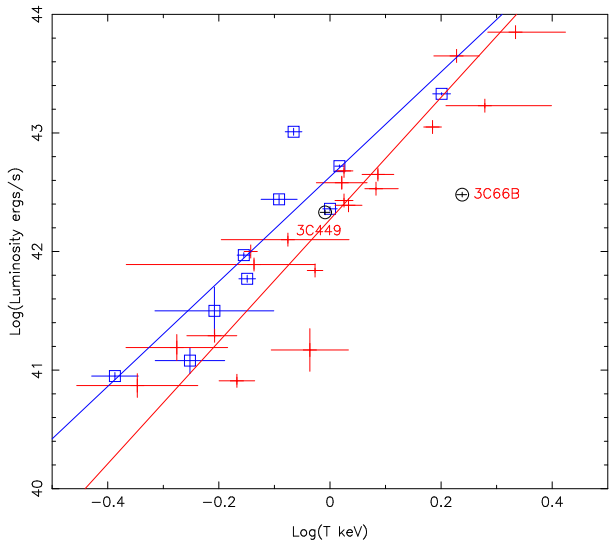}
\caption{Plot of luminosity vs. temperature for two samples of X-ray groups. On the left is shown sample 1 (indicated by hollow squares): the radio-quiet subsample extracted from Helsdon \& Ponman (2000), and sample 2 (indicated by + symbols), the combined radio-loud subsample of Helsdon \& Ponman (2000) and radio-galaxy sample of Worrall \& Birkinshaw (2000). Best-fitting luminosity-temperature relations for each sample are plotted, as decribed in the text, illustrating that the sample containing radio sources show an increased temperature for a given luminosity. 3C 66B and 3C 449 are included for comparison and marked with circles.}
\label{lt}
\end{center}
\end{figure}

We compared this radio-quiet subsample with a sample of radio-loud groups consisting of the groups from the Helsdon \& Ponman sample which were excluded from the radio-quiet subsample, combined with the sample of Worrall \& Birkinshaw (2000) (excluding 3C 449, as we have better constraints on its properties from this work) which contains low-power radio-galaxy groups from the B2 sample. Fig.~\ref{lt} shows our two samples, with 3C 449 and 3C 66B included for comparison. The best-fitting relation for each subsample is plotted. As there is a large intrinsic scatter in these relations, in order to determine realistic errors on parameters, we combined the measurement errors in temperature and luminosity with normally-distributed errors which are a small percentage of the temperature and luminosity. This extra error was increased until an acceptable fit was obtained for both samples; this required introducing an error of 0.005 $\times$ the logarithm of temperature and luminosity. The best-fitting relations are:
\begin{equation}
\log(L_{X}) = (4.42\pm0.51)\log(T) + (42.63\pm0.09)
\end{equation}
for the radio-quiet subsample (sample 1) and
\begin{equation}
\log(L_{X}) = (5.15\pm0.52)\log(T) + (42.27\pm0.07)
\end{equation}
for the combined radio-loud sample (sample 2), where $L_{X}$ is the bolometric X-ray luminosity in erg s$^{-1}$, and T is the temperature in keV. Both fits are consistent with a slope of $\sim$ 4.9, the value found by Helsdon \& Ponman; however, the intercepts for the two samples differ at greater than the 2$\sigma$ level. In addition we used a 1-D KS test to compare the temperature and luminosity distributions of sample 1 and sample 2. We find that the luminosity distributions of the samples are consistent with their being drawn from a single parent population, whereas the temperature distributions are different at the $\sim$ 90 per cent confidence level. 

Therefore, we find a small, but significant difference in the luminosity-temperature relation for groups containing radio galaxies, with higher temperatures found at a given luminosity. There are two possible interpretations for this result: either the radio galaxies are heating their environments, or they are reducing the luminosity of the groups. To test whether the second scenario could be true, we assumed that all of the material through which the radio source passes while expanding has been affected by the radio source and no longer emits in X-rays (e.g. it may have been lifted out to a distance where the subsequent decrease in density produces this effect). Modelling the expanding radio source as a cone, and comparing the square of number density integrated over the volume through which the radio source will have passed with the same integral for the entire group, we find that only 7 per cent of the luminosity of 3C 66B could have been removed in this way. Using the original group luminosity, a temperature of 0.92$\pm0.21$ keV is predicted from our best-fitting relation for radio-quiet groups. The effect will be at a similar level in other sources. Therefore we conclude that the radio galaxy could not have produced such a decrease in luminosity. 

For this reason, we interpret these results as providing indirect evidence for radio-source heating of group environments. This result shows that groups which contain detected radio sources are systematically hotter than those which do not, providing the first evidence that radio-source heating is occurring at a detectable level in the majority of such groups. It is apparent from Fig.~\ref{lt} that 3C 66B is much hotter than predicted by either of the two relations plotted. There are three contributing factors which could explain 3C 66B's exceptional behaviour. Firstly, it is more powerful than any of the sources in the radio-loud sample, and so would be expected to transfer more energy to the group gas. As described above, it is also likely to be comparatively old for an FR-I radio galaxy, so will have been putting energy into its environment for a longer period of time. Finally, the environment of this source is less massive than the environments of other large, powerful sources, such as 3C 31. It seems plausible that these three factors have combined to make 3C 66B particularly efficient at raising the temperature of its environment. 

\section{Conclusions}

Our images of 3C 66B and 3C 449 provide direct evidence that their environments are having a major effect on the evolution of the radio sources. In both cases large rounded lobes are associated with large amounts of X-ray-emitting gas, whereas narrower lobes are seen where there is no surrounding material. In the case of 3C 66B the presence of a blob of gas at the end of the eastern jet and bounded lobe is suggestive of an obstacle having a strong impact on the lobe morphology. 

We have revisited the luminosity-temperature relation for clusters and groups and find slightly different relations for samples of radio-quiet and radio-loud groups. We interpret this as indirect evidence that radio-source heating is occurring in the majority of groups containing radio galaxies, and explains the higher-than-predicted temperatures we observe for the environments of 3C 66B and 3C 449. Better constraints on the temperatures of radio-quiet and radio-loud groups are needed to confirm this result. We find that either these two sources are expanding supersonically, or they are significantly older than expected from spectral age estimates. Supersonic expansion is unlikely due to the appearances of the sources, therefore we favour the second explanation, which implies that the sources are old enough and sufficiently powerful to have heated their environments and produced the higher temperatures we observe.

The lobes of low-power radio galaxies are either out of equipartition or supported by a pressure contribution from particles which we cannot observe directly. The lack of detected X-ray emission from the radio lobes rules out entrained thermal material at the temperature of the environment. We can also constrain the magnetic field strength in the lobes because of the lack of X-ray inverse Compton emission, from which we conclude that if a departure from equipartition explains the apparent pressure imbalance, the lobes must be magnetically dominated.

\section*{Acknowledgments}

JHC thanks PPARC for a studentship. MJH thanks the Royal Society for a research fellowship.

\bsp

\label{lastpage}

\end{document}